\documentclass[aps,prd,a4paper,reprint,nofootinbib,superscriptaddress,floatfix,preprintnumbers]{revtex4-1}

\usepackage{hyperref}
\usepackage{amsmath}
\usepackage{amsfonts}
\usepackage{amssymb}
\usepackage{color}
\usepackage[utf8]{inputenc}
\usepackage{graphicx}
\usepackage{gensymb}
 
\usepackage{microtype}
\usepackage{siunitx}
\usepackage{soul}
\usepackage{rotating}

\usepackage{comment}

\newcommand{\Fermi}{\textit{Fermi}}
\newcommand{\SF}{\texttt{SkyFACT}}
\newcommand{\ST}{\texttt{ScienceTools}}

\usepackage{accents}
\newlength{\dhatheight}

\begin{document}

\begin{flushleft}
LAPTH-007/21
\end{flushleft}

\title{Gamma-ray image reconstruction of the Andromeda galaxy}
\date{\today}

\author{C\'eline Armand}\email{armand@lapth.cnrs.fr}
\affiliation{Univ.~Grenoble Alpes, USMB, CNRS, LAPTh, F-74000 Annecy, France}
\affiliation{Univ.~Grenoble Alpes, USMB, CNRS, LAPP, F-74000 Annecy, France}
\affiliation{Astronomy Department, University of Geneva, Chemin d'Ecogia 16, 1290 Versoix, Switzerland}
\author{Francesca Calore}
\affiliation{Univ.~Grenoble Alpes, USMB, CNRS, LAPTh, F-74000 Annecy, France}

\begin{abstract}
We analyze about 12 years of \Fermi-LAT data in the direction
of the Andromeda galaxy (M31). We robustly characterize its
spectral and morphological properties against systematic uncertainties related to 
the modeling of the Galactic diffuse emission. We perform this work by adapting and exploiting 
the potential of the \SF~adaptive template fitting algorithm. We reconstruct the {\it $\gamma$-ray image} of M31 in a template-independent way, and we show that flat spatial models are preferred by data, indicating an extension of the $\gamma$-ray emission of about $0.3-0.4\degree$ for the bulge of M31. This study also suggests that a second component, extending to at least $1\degree$, contributes to the observed total emission. We quantify systematic uncertainties related to mis-modeling of Galactic foreground emission at the level of 2.9\%.

\end{abstract}

\maketitle
\section{Introduction}\label{intro}

Distant galaxies constitute promising targets to understand astrophysical processes occurring inside such systems. Over the past decade, several star-forming galaxies have been detected in $\gamma$ rays by \Fermi-LAT, among which are the Large Magellanic Cloud, the Small Magellanic Cloud, Andromeda galaxy, M82, NGC 253, NGC 2146, and Arp 220. The $\gamma$-ray emission of these distant galaxies is produced by the interactions of the large-scale population of cosmic rays with the interstellar medium and by high-energy sources such as supernova remnants and pulsars therein~\cite{Ackermann:2017nya}. The study of external Milky Way-like galaxies can provide an independent and outside perspective of the $\gamma$-ray astrophysical processes which are in place in our own Galaxy as well~\cite{DiMauro:2019frs}. Therefore, it would bring out additional and complementary knowledge to our understanding of the Milky Way, especially if we can discriminate individual contributions to the $\gamma$-ray emission of such extragalactic systems. Necessary (but not sufficient) condition to this end is that an extended emission signal is significantly detected in $\gamma$ rays from the external galaxy.
With this ultimate goal in mind, in the present work, we focus on our nearest galaxy neighbor, the Andromeda spiral galaxy (M31), located at a distance of approximately $785 \pm 25$~kpc\footnote{Several measurements of the distance have been performed 
ranging from $765\pm 28 $~kpc~\cite{Riess_2012} to  $790\pm 45$~kpc~\cite{Joshi:2002uf}. Most of them are found around 775-785~kpc~\cite{Stanek:1998cu, Conn_2012, Durrell:2001zt, Holland:1998br, Ribas:2005uw}. In our study, we adopt a distance of $785 \pm 25$~kpc~\cite{McConnachie:2004dv} to be consistent with previous works on the subject that refer to this value.}~\cite{McConnachie:2004dv} at Galactic coordinates
($l=121.285\degree$, $b=-21.604\degree$)~\cite{Fermi-LAT:2019yla}.

Its close proximity allows us to optically resolve its stellar disk and bulge as two separate components. This distinction is not possible in our Galaxy as the bulge is obscured by the bright emission of the disk~\cite{Ackermann:2017nya}. 
M31 spans $3.2\degree\times1\degree$ on the sky~\cite{DiMauro:2019frs, Feng:2018vsl}.
This spiral galaxy has a total mass of $(0.7-2.1)\times10^{12}~M_{\odot} $~\cite{Tempel:2007eu, Fardal:2013asa, Widrow:2003yi, Evans:2000wd, Evans:2000cm, Seigar:2006ia, Tamm_2012, Corbelli_2010}. The stellar component accounts for $(10-15)\times 10^{10}~M_{\odot}$ of which 30\% is in the bulge and 56\% is in the disk~\cite{Tamm_2012}.
Furthermore, disk galaxies are typically surrounded by a large cosmic-ray halo extending up to a few hundreds of kpc~\cite{Feldmann_2012}, as well as by a circumgalactic medium made mostly by ionised hydrogen which can extend up to the virial radius~\cite{Lehner_2015}.

As for $\gamma$-ray studies, M31 has the advantage of lying at high Galactic latitudes, away from the Galactic plane which makes it less polluted by the diffuse $\gamma$-ray foreground emission of the Milky Way~\cite{Li:2013qya}.
M31 has been the object of several dedicated $\gamma$-ray analyses 
aimed at its spectral and morphological characterization. 
The first study in this direction was performed by the \Fermi-LAT collaboration who analyzed about two years of \Fermi-LAT data, detecting M31 as a point-like source
with a $5.3\sigma$ significance and founding marginal evidence ($1.8\sigma$) for its spatial extension~\cite{Fermi_M31_M33}. 
A seven-year data analysis showed evidence for point-like source detection at $\sim10\sigma$, and a $4\sigma$ preference for extended emission, compatible with either a Gaussian distribution of width $0.23\degree \pm 0.08$ or a uniform disk of radius $0.38\degree \pm 0.05$. 
Several studies tested the extension of M31 in $\gamma$ rays 
using various models for its morphology~\cite{Li:2013qya, Pshirkov:2016qhu, Feng:2018vsl, Karwin:2019jpy, DiMauro:2019frs}, mostly confirming the preference for a centrally concentrated emission. Additionally,~\cite{Pshirkov:2016qhu} found weak evidence for the presence of ``Fermi bubble''-like structures perpendicularly to M31's galactic plane, while~\cite{Karwin:2019jpy, Do:2020xli} showed the existence of an extended excess up to $120-200$ kpc ($8-15\degree$) away from the center of M31.

Interestingly, no evidence for $\gamma$-ray emission correlated with M31's gas-rich regions or star formation activity was found. On the contrary, a mild correlation with infrared stellar templates emerged, hinting to a possible origin from old stars~\cite{Ackermann:2017nya}.
Such a centrally concentrated signal, perhaps associated with an old stellar population in M31, 
was suggestive of another longstanding excess in $\gamma$-ray astrophysics, the so-called \Fermi~Galactic center GeV excess~\cite{Goodenough:2009gk, Hooper:2010mq, Hooper:2011ti, Hooper:2013rwa, Daylan:2014rsa, TheFermi-LAT:2015kwa, Calore:2014xka, vitale2009indirect, Abazajian:2012pn, Boyarsky:2010dr, Huang:2013pda}.
A possible connection between the two signals was explored by~\cite{Eckner:2017oul}, which discussed the possibility of a common origin from primordial and dynamically formed millisecond pulsars.
Possibly, M31 extended signal can also be compatible with dark matter annihilation from an adiabatically contracted dark matter density profile~\cite{DiMauro:2019frs}.

Despite the growing evidence for M31 $\gamma$-ray spatial extension, the main limitation 
of current analyses remains the evaluation of the systematic uncertainties related to the contamination of the Galactic diffuse emission, as discussed also in~\cite{Karwin:2019jpy,DiMauro:2019frs}.

In what follows, we use the \SF~adaptive template fitting algorithm~\cite{Storm:2017arh} to characterize the $\gamma$-ray emission in the direction of M31.
Our goal is twofold: first, we aim to robustly detect M31 as an extended source against foreground model systematic uncertainties, to characterize its spectral and spatial distributions, and to test the presence of multiple $\gamma$-ray emission components. To this end, we will for the first time include in the fit M31 morphology templates tracing its stellar distribution from~\cite{Tamm_2012}. We willl perform spectral and spatial fits of several models to the data and make the first proper comparison of our non-nested models.
Secondly, we seek to reconstruct M31's morphology in a fully data-driven way based on the image reconstruction feature of the \SF~code. 
This work represents the first study where a ``template-independent'' approach is applied to the analysis of an extended $\gamma$-ray signal. This image reconstruction feature allows us to build the galaxy intensity profile and derive its parametrization.

The paper is organized as follows: In Sec.~\ref{data}, we present the data selection and the fitting algorithm. Section~\ref{Modeling_gamma_ray_emission} briefly describes how we model the different emission templates required in the analysis. 
Section~\ref{sec:std} illustrates the results obtained in the case of standard template fitting techniques and it is meant to reproduce and strengthen literature results. In Sec.~\ref{sec:semiATF}, we exploit the innovative features of \SF~and demonstrate that the evidence for extension is robust against foreground contamination.  In Sec.~\ref{sec:atf}, we model-independently reconstruct the morphology of M31 and provide its intensity profile. We finally discuss the results and conclude on this work in Sec.~\ref{sec:conclusion}.

\section{Data selection and fitting algorithm}\label{data}

\subsection{Data selection}

We use 624 weeks (12 years) of \Fermi-LAT data from August 4th, 2008 to July 16th, 2020, collected 
from a $10\degree \times 10\degree$ region of interest (ROI) centered at ($l=121\degree$,  $b=-21\degree$).
We 
select Pass 8 SOURCE class events with Point Spread Function (PSF) PSF0+PSF1+PSF2+PSF3 type and use the corresponding instrument response functions (IRFs) P8R3 SOURCE V2. We apply a cut on maximum apparent zenith angle ($z_{\rm{max}} = 105\degree$), and recommended data-quality filters
(DATA\_QUAL$>$0) \& (LAT\_CONFIG==1). 
We analyze data in the energy range 300~MeV to 100~GeV, logarithmically spaced into 22 bins.
Table~\ref{Criteria_data} summarizes the data selection criteria. This work makes use of the \Fermi~\ST~v11r5p3 software package\footnote{\url{http://fermi.gsfc.nasa.gov/ssc/data/analysis/}}. We spatially split the data of the ROI into $200 \times 200$ angular bins of size $0.05\degree$ in cartesian sky projection.

\begin{table}[ht!]
\centering
{
\footnotesize
\begin{tabular}{c|c}
\hline
 & Event selection criteria\\
\hline \hline
Event class & Pass 8 SOURCE  \\

Event type &  PSF0+PSF1+PSF2+PSF3 \\

IRFs &  P8R3 SOURCE V2 \\

$z_{\rm{max}}$ &  105\degree \\

ROI size & $10\degree \times 10\degree$\\

ROI center & $l=121\degree$,  $b=-21\degree$\\

Pixel resolution & 0.05\degree\\

Pixel binning & 200 $\times$ 200 pixels\\

Sky projection & Cartesian ``CAR''\\

Energy range & 300~MeV to 100~GeV\\

Energy binning & 22 bins \\

Filters & (DATA\_QUAL$>$0) \& (LAT\_CONFIG==1)\\
\hline
\end{tabular}
}
\caption{Summary of the selection criteria applied to the data using the \Fermi~\ST.}
\label{Criteria_data}
\end{table}

\subsection{\SF}
To perform the $\gamma$-ray analysis of M31 we use \SF~\cite{Storm:2017arh}, a code which combines a hybrid approach template fitting and image reconstruction for studying and decomposing the $\gamma$-ray emission. 
\SF~offers the advantage of accounting for expected spatial and spectral uncertainties in each model emission component of the fit, contrary to standard template fitting methods
where the spatial distributions of model components are fixed. 
The algorithm is based on penalized maximum likelihood regression, with additional nuisance parameters that account for uncertainties modeling the imperfections of the model templates. More details on the analysis code can be found in Appendix~\ref{App:SF}.

The detection of the source and the evidence for its extension are probed through a likelihood ratio (LR) statistical test, denoted TS, between our various models including M31 and our model without the source. The TS results are then interpreted in terms of standard deviations $\sigma$ using the Chernoff theorem, describing a mixture between a $\chi^2$ and a Dirac function. A detailed description of the statistical analysis framework can be found in Appendix~\ref{stats}.

\section{Modeling the $\gamma$-ray emission}\label{Modeling_gamma_ray_emission}

In our model the $\gamma$-ray emission in the ROI is made up by the following contributions: the Galactic interstellar emission (IEM), the Isotropic Gamma Ray Background (IGRB), 13 Point-like Sources (PS) as reported in the 4FGL~\cite{Fermi-LAT:2019yla}, and M31. 
No extended source is present in the chosen ROI.
We describe below how we model the individual templates (spectral and spatial), used by \SF~in the fit of the data.

\subsection{IEM, IGRB and PS modeling}
The Galactic IEM represents the foreground emission from the Milky Way, and includes  
the contribution from Inverse Compton, $\pi^0$ decay, and Bremstrahlung $\gamma$-ray production mechanisms. 
We adapt the Galactic IEM \texttt{gll\char`_iem\char`_v06.fits}\footnote{We voluntarily use an older version of the IEM template model (\texttt{v06}) since the latest one (\texttt{v07}) may include some emission in the direction of M31, and due to M31 gas not subtracted from that IEM model.} provided by~\cite{Fermi_model_bkg} to our energy and spatial binnings, and normalize the spatial template by the sum of all pixels for each energy bin. The normalization of each energy bin corresponds to the spectral profile of this diffuse component.

As spectral template of the IGRB component, we use  \texttt{P8R3\char`_SOURCE\char`_V2.txt}~\cite{Fermi_model_bkg} associated to \texttt{gll\char`_iem\char`_v06.fits}. We produce a flat template to model the spatial part, as the IGRB is assumed to be isotropic after exposure correction of the LAT~\cite{DiMauro:2015tfa}.

PS best-fit spectra are extracted from the 4FGL catalog, \texttt{gll\_psc\_v22.fits}, of \Fermi-LAT sources~\cite{Fermi-LAT:2019yla}, and are used as input spectra for the \SF~fit.

\subsection{M31 modeling}\label{sec:M31_templates_descrip}
As for M31 galaxy, we model its $\gamma$-ray emission using various templates motivated either by previous works~\cite{Feng:2018vsl, DiMauro:2019frs, Li:2013qya}  or by studies of the stellar distribution of the galaxy~\cite{Tamm_2012}, with the final goal of understanding which M31 morphology is preferred by data. 

As for the morphology (\textit{i.e.}~spatial component) of M31, we test the following distributions: 

\paragraph{Gaussian spatial profile.}
We build a first set of spatial templates for M31 using a Gaussian function whose width $\sigma$ defines the 68\% containment extension angle.
We produce Gaussian templates of width ranging from $0.001\degree$ (PS limit) to $1\degree$ around the position of M31.

\paragraph{Uniform spatial profile.}
A second set of spatial templates is constructed with uniform disks of radii extending from $0.025\degree$ (PS limit) to $1\degree$. 

\paragraph{Einasto spatial profile.}
Finally, we model M31 as three distinct stellar components for the nucleus, bulge, and disk, using three ellipses inclined at $77.5\degree$~\cite{MAJun:1420}. 
The density distribution of each of them follows the Einasto law whose parameters are defined in~\cite{Tamm_2012} and reads: 

\begin{equation}
\rho_{\rm{Ein}} (r) = \rho_c \exp \left\{ -d_N \left[ \left(\frac{r}{r_c} \right)^{1/N} - 1 \right]   \right\},
\end{equation}
where $d_N$ is a function of the coefficient $N$ defined as $d_N \approx 3N -1/3 + 0.0079/N$ for $N \geq 0.5$. The parameter $\rho_c$ corresponds to the density at a distance $r_c$ which encloses half of the total mass of the component.
The variable $r$ is the distance of a given pixel from the center of M31 expressed in terms of line of sight $s$ as:
\begin{equation}
r(s,b,l) =  (d^2 + s^2 - 2ds \cos{b} \cos{l})^{1/2}
\label{distance_r}
\end{equation}
where $(b,l)$ are the relative galactic coordinates to the center of M31 and $d$ is the distance between the observer and M31.

We integrate the Einasto profile over the line of sight towards each pixel $p$ of the ROI in order to obtain a spatial template. The integration is performed according to:
\begin{equation}
T_p(b,l) = \int_s \rho_{\rm{Ein}}(r(s,b,l))  ds \, ,
\end{equation}
whose upper and lower limits are derived by solving Eq.~\ref{distance_r} for $s$ with $r=R_{\rm{vir}}$ where $R_{\rm{vir}}$ is the virial radius, set to 213 kpc~\cite{Tamm_2012}.
The two solutions obtained for $s$ are:
\begin{equation}
s_{\rm{max/min}} = d\cos b \cos l \pm \sqrt{ d^2 [(\cos b \cos l)^2 - 1]  +  R^2_{\rm{vir}} }
\end{equation}

We consider three sub-components to model M31, \textit{i.e.} the nucleus, the bulge, and the disk, each defined by an ellipse of ratio $q$ and an Einasto parametrization. 
We note that 68\% of the total emission expected for each component is contained in an ellipse of major (minor) axis 2.15$\degree$ (0.37$\degree$) in the case of the disk, and 0.26$\degree$ (0.18$\degree$) in the case of the bulge. For the nucleus template, instead, about 95\% of the total emission comes from the central pixel.
Table~\ref{ellipse_einasto} presents the parameters $\rho_c$, $r_c$, $N$, $d_N$, and $q$ characterizing the spatial template of each sub-component.
In what follows, we will consider these templates individually or combined in order to model the emission of M31. 

\begin{table}[ht!]

\centering{\textbf{\sffamily\bfseries  Ellipse parameters and Einasto templates }}
\vspace{0.2cm}

\centering
{
\footnotesize
\begin{tabular}{c|c|c|c|c||c}
\hline
Component &$\rho_c$ &$r_c$ & $N$ & $d_N$ & $q$ \\
\hline \hline
Nucleus &   1.713   & 0.0234  & 4.0 & 11.67 & 0.99  \\
\hline
Bulge &   $ 0.920 $  & 1.155   & 2.7 & 7.77& 0.72 \\
\hline
Disk &   $ 0.013 $   & 10.67   & 1.2 & 3.27&  0.17\\
\hline
\end{tabular}
}
\caption{Parameters of the Einasto templates where $\rho_c$ [$M_{\odot}.\rm{pc}^{-3}$] is the density, $r_c$ [kpc] is the radius enclosing half of the total mass of the sub-component, $d_N$ is a function of the component $N$, and $q$ the ratio between the major and minor axes of the ellipse. These values are taken from~\cite{Tamm_2012}.}
\label{ellipse_einasto}
\end{table}

As for the M31 spectral template, we use the M31 spectral energy distribution  
as provided in the 4FGL catalog, and modeled by a power law with exponential cutoff.

\section{Standard template fitting}\label{sec:std}

Although \SF~was developed for simultaneous spatial and spectral template fitting, it can also be used to perform traditional or standard template fits (STF). 
The analysis within this setup allows us to directly compare with the previous literature, as well as with the other \SF~runs we will explore in the following sections. 
For this purpose, we set all spatial hyper-parameters 
so that the spatial templates are kept fixed to the input models in the fit, for all components. By contrast, the spectral parameters are left completely free to vary, \textit{i.e.}~bin-by-bin energy independent fit.

\subsection{Detection of M31 as PS}

\begin{figure*}[p!]
\centering{
\includegraphics[angle=0,width=0.45\textwidth]{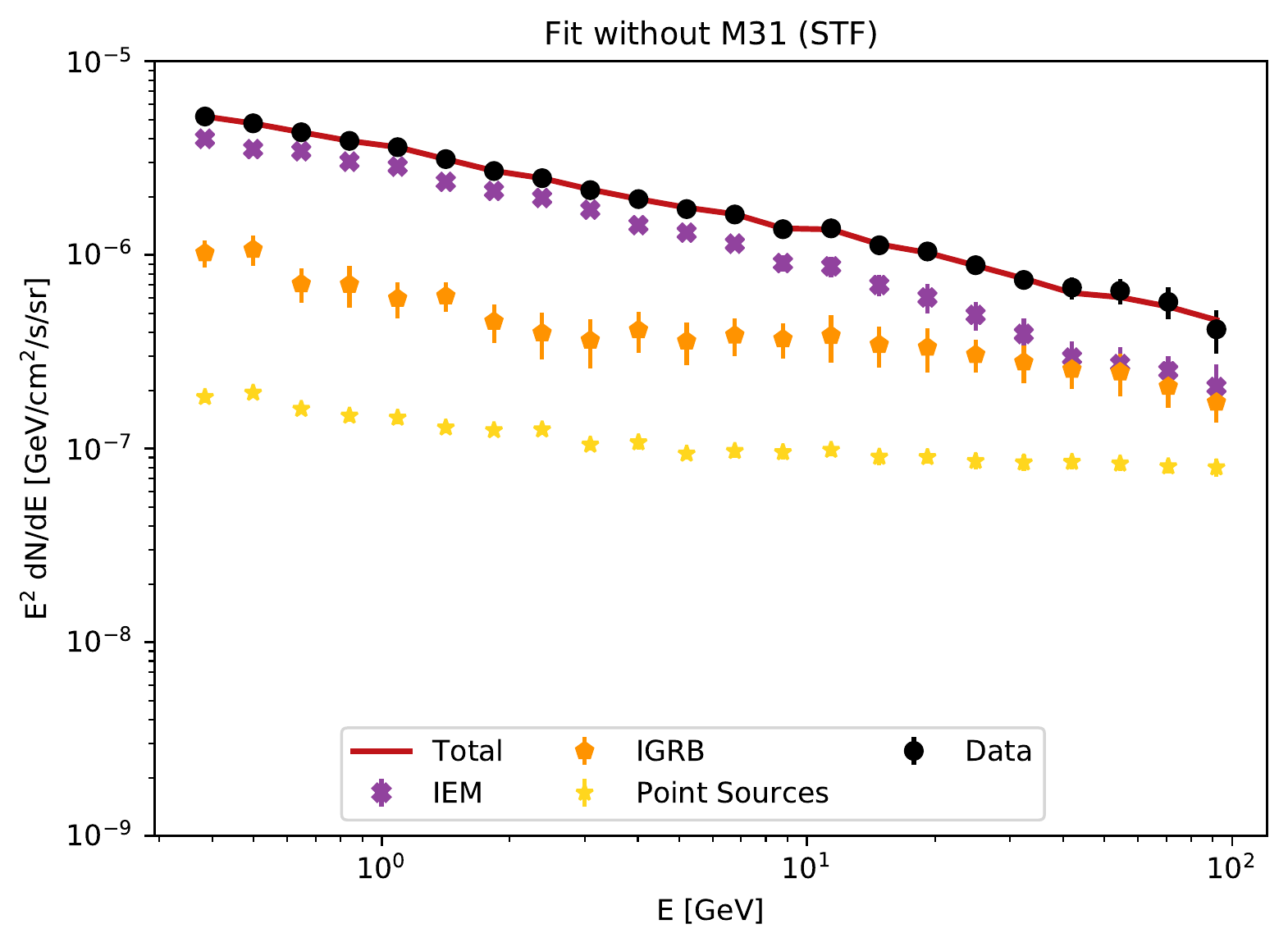} 
\includegraphics[angle=0,width=0.45\textwidth]{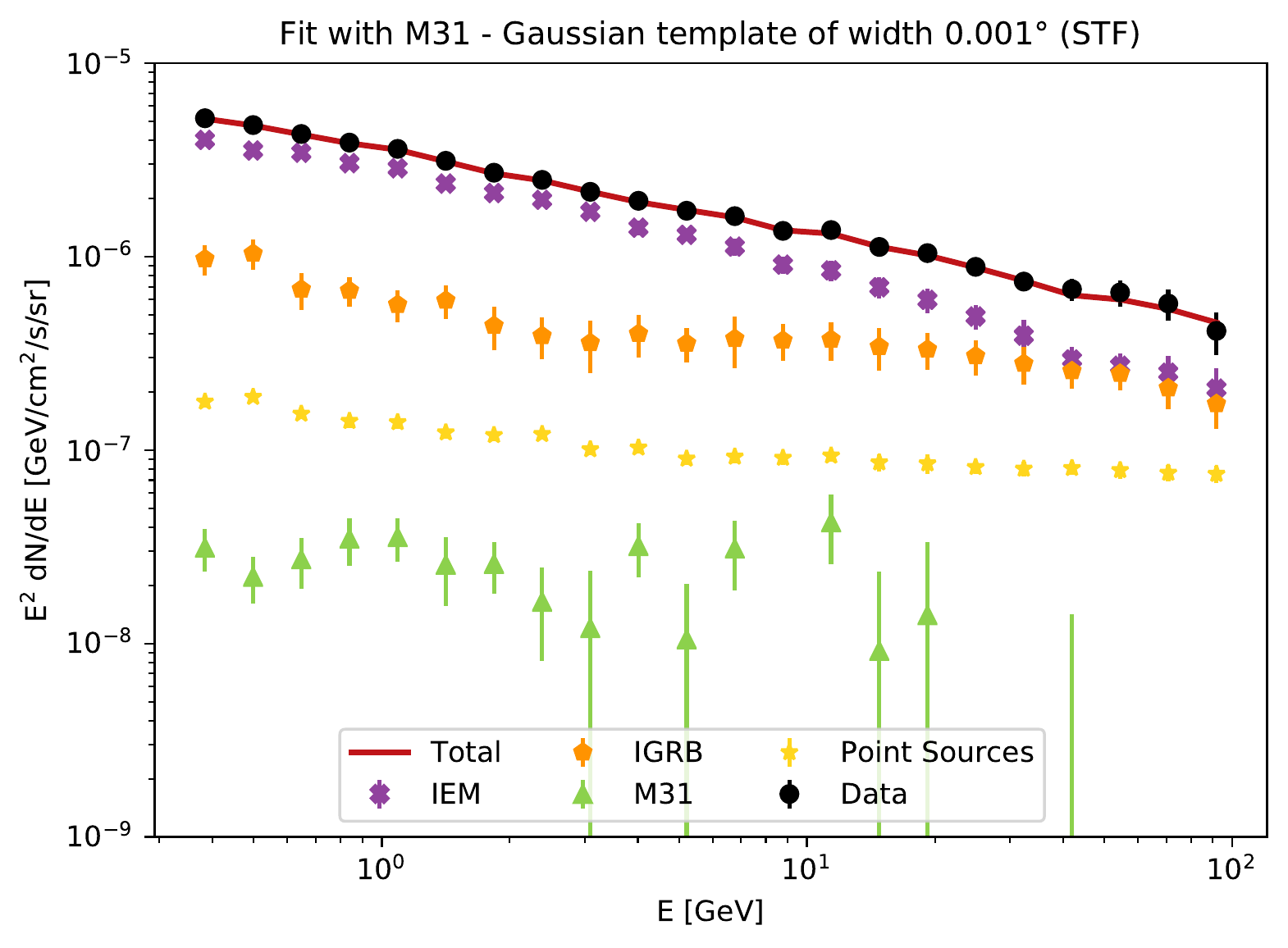}}
\caption{Spectra of each component for the fit without M31 \textbf{(left)} and with M31 modeled as a 2D Gaussian of width $0.001\degree$ (PS-limit) \textbf{(right)}.}
 \label{fig:std_spec} 
\end{figure*}

\begin{figure*}[p!]
\centering{
\includegraphics[angle=0,width=0.45\textwidth]{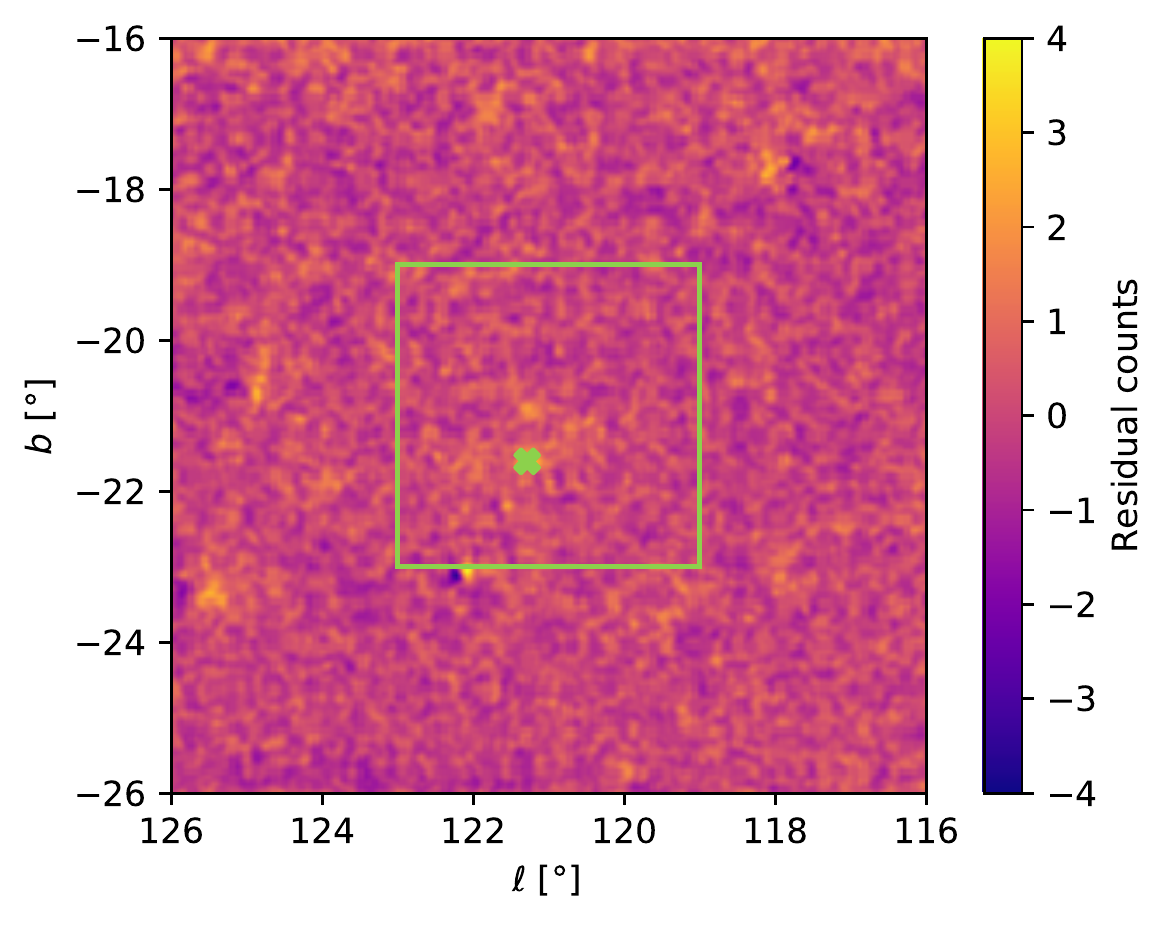}
\includegraphics[angle=0,width=0.45\textwidth]{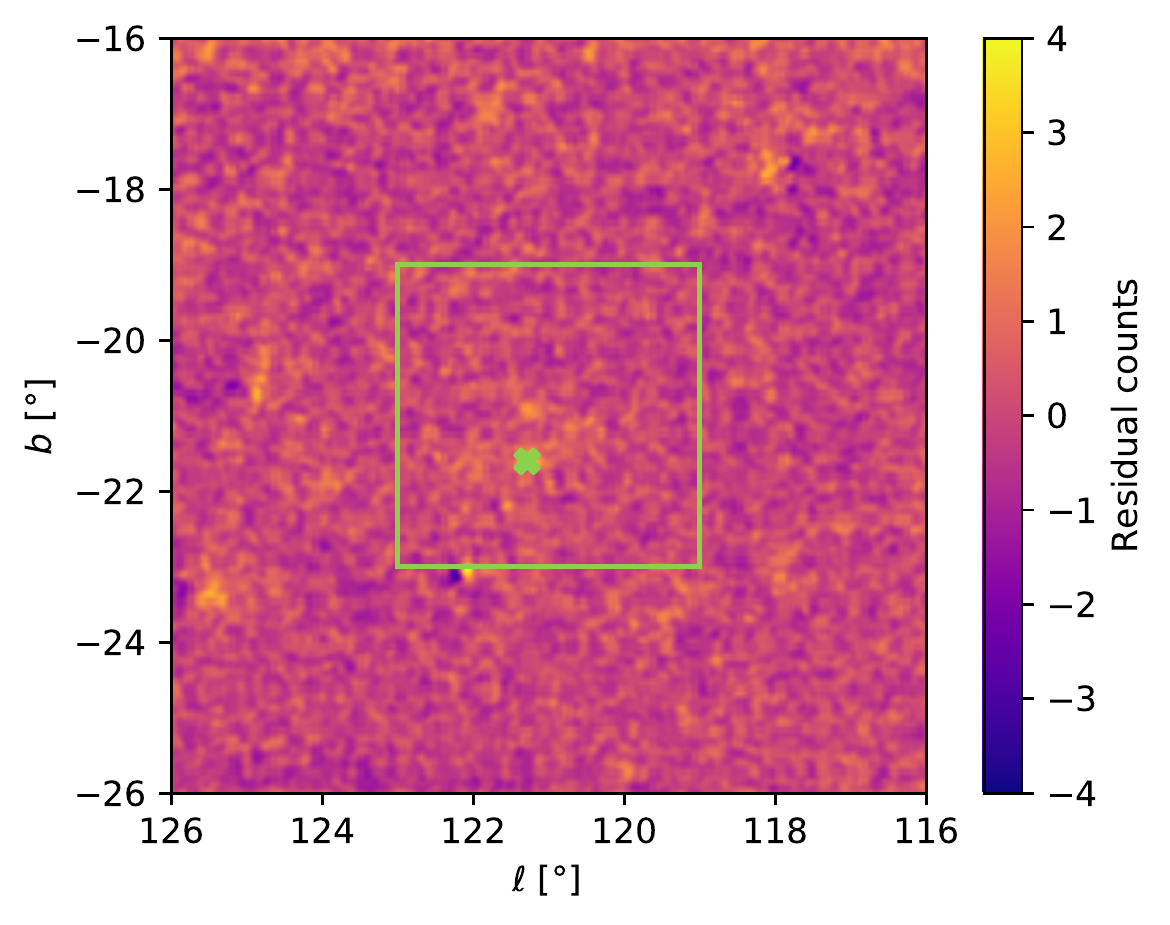}
\includegraphics[angle=0,width=0.45\textwidth]{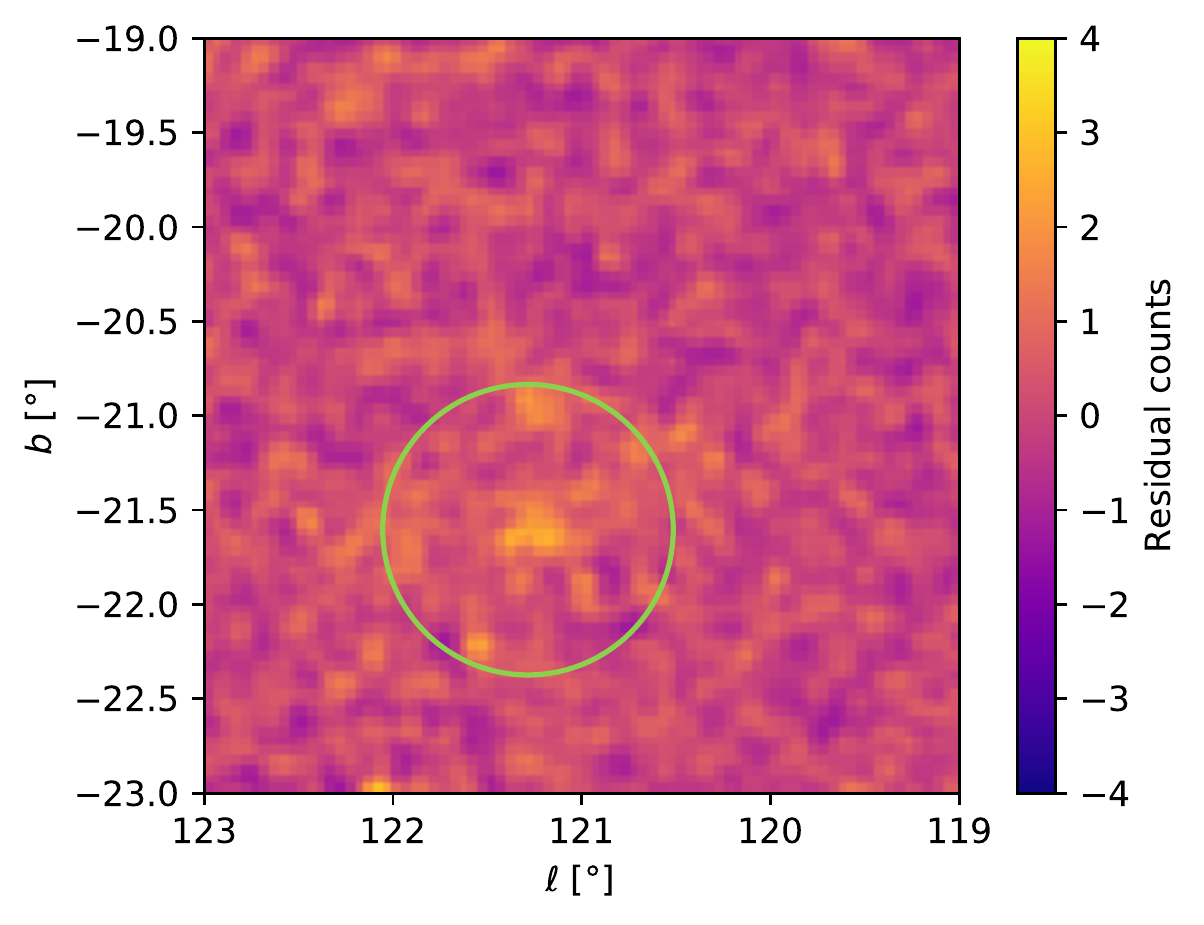}
\includegraphics[angle=0,width=0.45\textwidth]{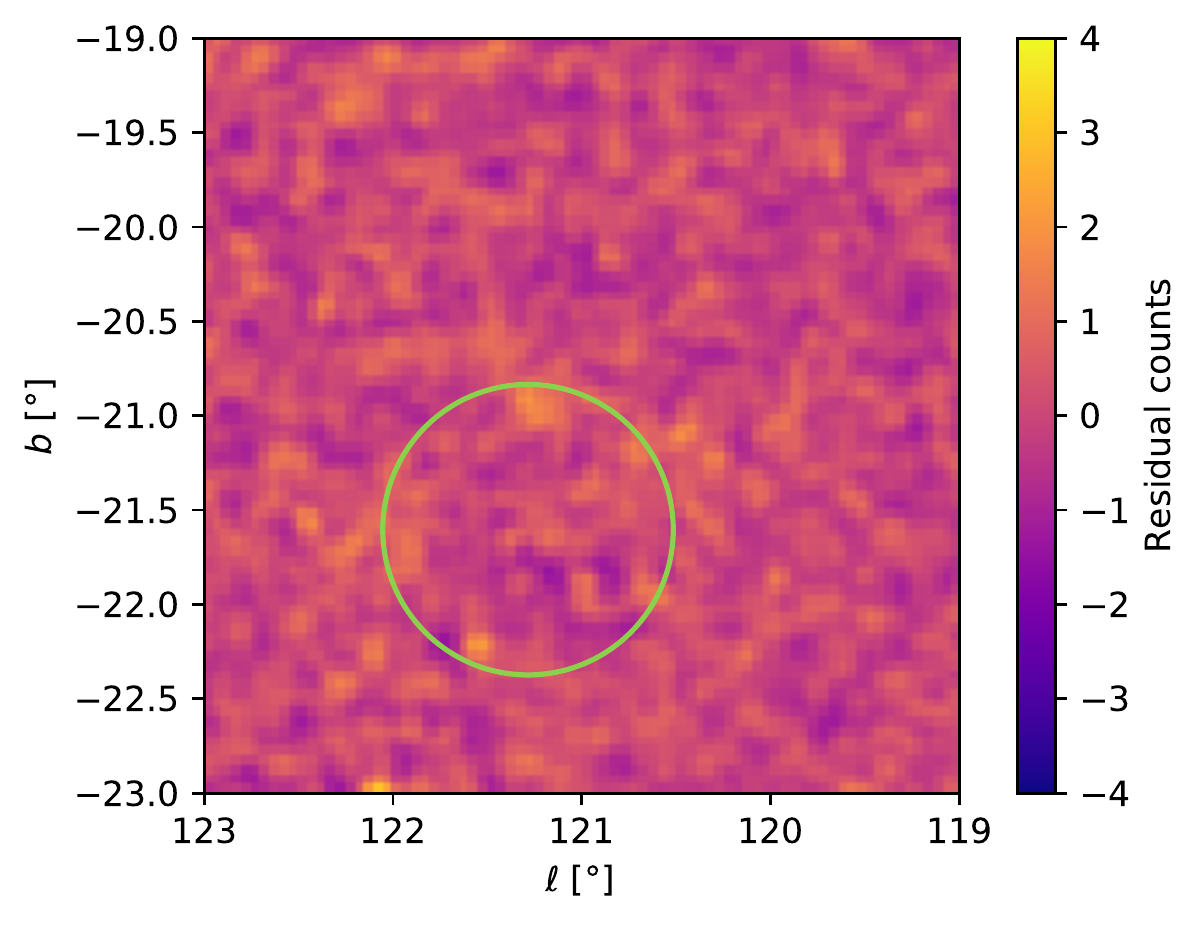} }
\caption{
Residual maps for the fits without \textbf{(left column)} and with \textbf{(right column)} the M31 component. The residual maps show the difference between the data and the theoretical model.
The top row shows the residuals of the whole ROI where the center of M31 is marked by a green cross, while the bottom row is a zoom-in of the area within the green square. 
The zoom-in residual maps show that the model including the additional M31 component \textbf{(bottom right)} captures some of the $\gamma$-ray residuals clearly present when M31 is not included in the model (circled in green) \textbf{(bottom left)}.}
 \label{fig:std_res} 
\end{figure*}

We first perform STF to probe the evidence for M31 in \Fermi-LAT data in the energy range between 300~MeV and 100~GeV. 
We compare the fit of a model that only contains IEM, IGRB, and the 13 PS, \textit{i.e.} our \textit{STD baseline model}, to the fit of a second model including an additional component modeling M31.
For our PS-limit Gaussian spatial template ($0.001\degree$ width)
and input spectrum taken from the 4FGL, we compute the significance of an additional source in a nested model with bin-by-bin free spectral parameters according to the Chernoff theorem (see Appendix~\ref{stats}).
M31 is significantly detected in this $\gamma$-ray energy range with a $7.6\sigma$ significance.
Figures~\ref{fig:std_spec} and~\ref{fig:std_res} show respectively the flux of each component of the model fitted to the data, and the residuals in the absence of M31 compared to the fit including M31. 
We found a reconstructed spectral flux for M31 consistent with the one reported in the 4FGL catalog and parametrized by a power law with exponential cut-off.
We note that the lack of statistics generates very large uncertainties on the reconstruction of the M31 flux in the highest energy bins.
From the residuals map, we can see that the addition of a source at the M31 position captures some of the residuals left out by the model without M31.

\subsection{Evidence for the extension of M31}
\begin{figure}[h!]
\centering
\includegraphics[width=0.45\textwidth]{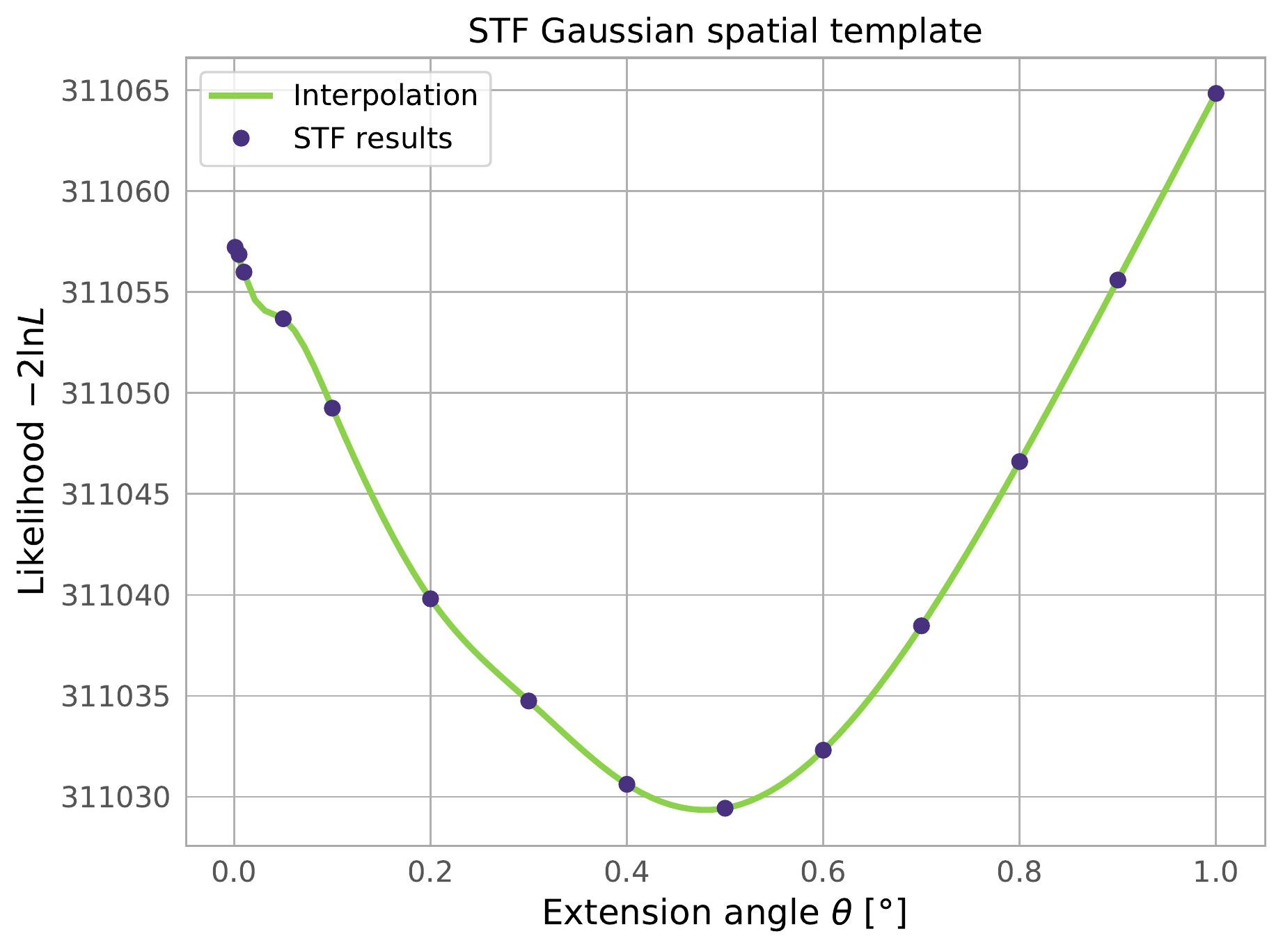}
\caption{Log-likelihood, $-2\ln \mathcal{L}$, profile from the scan over the extension angle $\theta$ (dots) of a Gaussian spatial template. The solid green line corresponds to the interpolation from which we estimate the best-fit extension angle and its statistical uncertainties.} 
\label{fig:STD_bestext} 
\end{figure}

In order to (i) demonstrate the preference for an extended emission morphology and 
(ii) measure the extension of the signal, 
we perform a scan over the 68\% extension angle $\theta$ of the Gaussian spatial profile\footnote{We will discuss below the extension measurement for other types of template.}.
We therefore obtain the log-likelihood ($- 2 \ln \mathcal{L}$) profile as a function of $\theta$, as shown in Fig.~\ref{fig:STD_bestext}.
The log-likelihood profile is minimized by non-zero extension angles, indicating a preference for an extended $\gamma$-ray emission for the M31 model component. 
The highest detection significance of the source is $\sigma = 9.2$, and it is associated to the Gaussian template of width $\theta=0.5\degree$.

In order to estimate the best-fit extension angle, 
we fit the log-likelihood profile with a cubic spline (green line in Fig.~\ref{fig:STD_bestext}). The minimum of the curve corresponds to 
 $\theta = 0.48\degree \pm{0.07}$. 
The uncertainties on the extension are computed by determining the two values of $\theta$ at which:
\begin{equation}
-2 \ln {\mathcal L}(\theta \pm \Delta \theta ) = -2 \ln\mathcal{L}(\hat{\theta}) + \left(-2 \Delta \ln \mathcal{L(\theta)} \right)
\end{equation}
where $2 \Delta \ln \mathcal{L(\theta)} = 1.00$ for a coverage probability of $68.27\%$ in the case of one parameter \cite{PDG_2020}. 
The $1\sigma$ statistical uncertainties of each side of the distribution are then derived by the following subtractions:
\begin{equation}
\sigma_{\theta}^{+} = \theta^+ - \hat{\theta}; \quad \quad \quad \sigma_{\theta}^{-} = \theta^- - \hat{\theta}.
\end{equation}
This result is compatible with the value of \cite{DiMauro:2019frs} who found a size of $0.42\degree \pm{0 .10}$ using a Gaussian template.

\subsection{Systematics due to PSF mis-modeling}
We quantify the impact of systematic uncertainties on the angular extension $\theta$
due to the possible mis-modeling of the PSF.
To this end, we build two alternative PSF kernels based on the prescription of~\cite{Biteau:2018tmv}, where the average PSF, $P(\theta, E)$, is modified according to:

\begin{equation}
\left\{\begin{array}{ll} P_{\rm{min}}(\theta;E)=P(\theta \times [1+f(E)] ; E)(1+f(E))^2 &\\ \\
P_{\rm{max}}(\theta;E)=P(\theta \times [1+f(E)]^{-1} ; E)(1+f(E))^{-2} &,
 \end{array} \right. 
\end{equation}
where $\theta$ is the reconstruction angle and $E$ is the energy. The function $f(E)$ corresponds to the scaling function for the relative PSF uncertainty depending on the energy and is defined by the following:

\begin{equation}
f(E) = \left\{
\begin{aligned}[lcl] 0.05 & & E \leq 10~\rm{GeV}\\
 0.05 + 0.1\log_{10} (E/10~\rm{GeV}) &&  E > 10~\rm{GeV}
 \end{aligned} \right.,
\end{equation}
where the uncertainty is constant at $5\%$ below $10$~GeV and increases up to $25\%$ at an energy of 1~TeV.
Using these two bracketing models of the PSF, we scan over the angular extension $\theta$ and we find the corresponding best-fit angular extensions: $\theta_{\rm{P}_{\rm{min}}} = 0.48\degree \substack{+0.07\\-0.06}$ and $\theta_{\rm{P}_{\rm{max}}} = 0.47\degree \substack{+0.07\\-0.07}$.
To estimate the systematic uncertainty on the quantity of interest, $\theta$, 
we follow Eq.~4 of~\cite{Biteau:2018tmv}, and compute the dispersion between the nominal value of $\theta$ and the values obtained with the two bracketing PSF kernels $P_{\rm{min}}(\theta;E)$ and $P_{\rm{max}}(\theta;E)$.
We find the systematic uncertainty on $\theta$ 
due to PSF mis-modeling to be 1.5\%.

\section{Semi-Adaptative Template Fitting}\label{sec:semiATF}
After having obtained results for the STF setup, we use \texttt{SkyFACT} to test systematic uncertainties
related to the uncertainties on the foreground emission template.
In particular, by allowing some more freedom on the spatial hyper-parameters of the IEM diffuse component (in addition to the full freedom on the spectral template already tested by the STF approach), we can test (i) whether or not the extension of the source is still preferred against uncertainties on the IEM template, and (ii) the robustness of the reconstructed spectra of M31 and IEM components against the freedom given to the baseline parameters.

In what follows, we refer to this setup as 
Semi-Adaptative Template Fitting (sATF).
We define three configurations of baseline models allowing a 30\%, 50\%, and 100\% variation of the IEM spatial parameters. 
We test the evidence for M31 and measure its extension on top of each of these baseline models and for different M31 spatial profiles: Gaussian, uniform, and Einasto templates.

\subsection{sATF with Gaussian spatial templates}
We perform three scans over $\theta$, one for each new baseline model, and we derive the best-fit angular extension from the cubic spline interpolation of the log-likelihood profile.  
The three IEM configurations give compatible values with best-fit extension angles of 
$\theta = 0.48\degree\pm0.07$ (IEM 30\%), 
$\theta = 0.48\degree\substack{+0.07 \\ -0.08}$ (IEM 50\%), and 
$\theta = 0.45\degree\substack{+0.09 \\ -0.10}$ (IEM 100\%).
These results are also in agreement with the STF results of Sec.~\ref{sec:std}. 
We evaluate the systematic uncertainty on $\theta$ due to the IEM mis-modeling to be 2.9\%, by computing the dispersion among the nominal value from the STF scan (no freedom allowed on the IEM component) and the best-fit extensions obtained with the three sATF configurations, with Eq.~4 of~\cite{Biteau:2018tmv}.

\begin{figure}[h!]
\centering
\includegraphics[width = 0.45\textwidth]{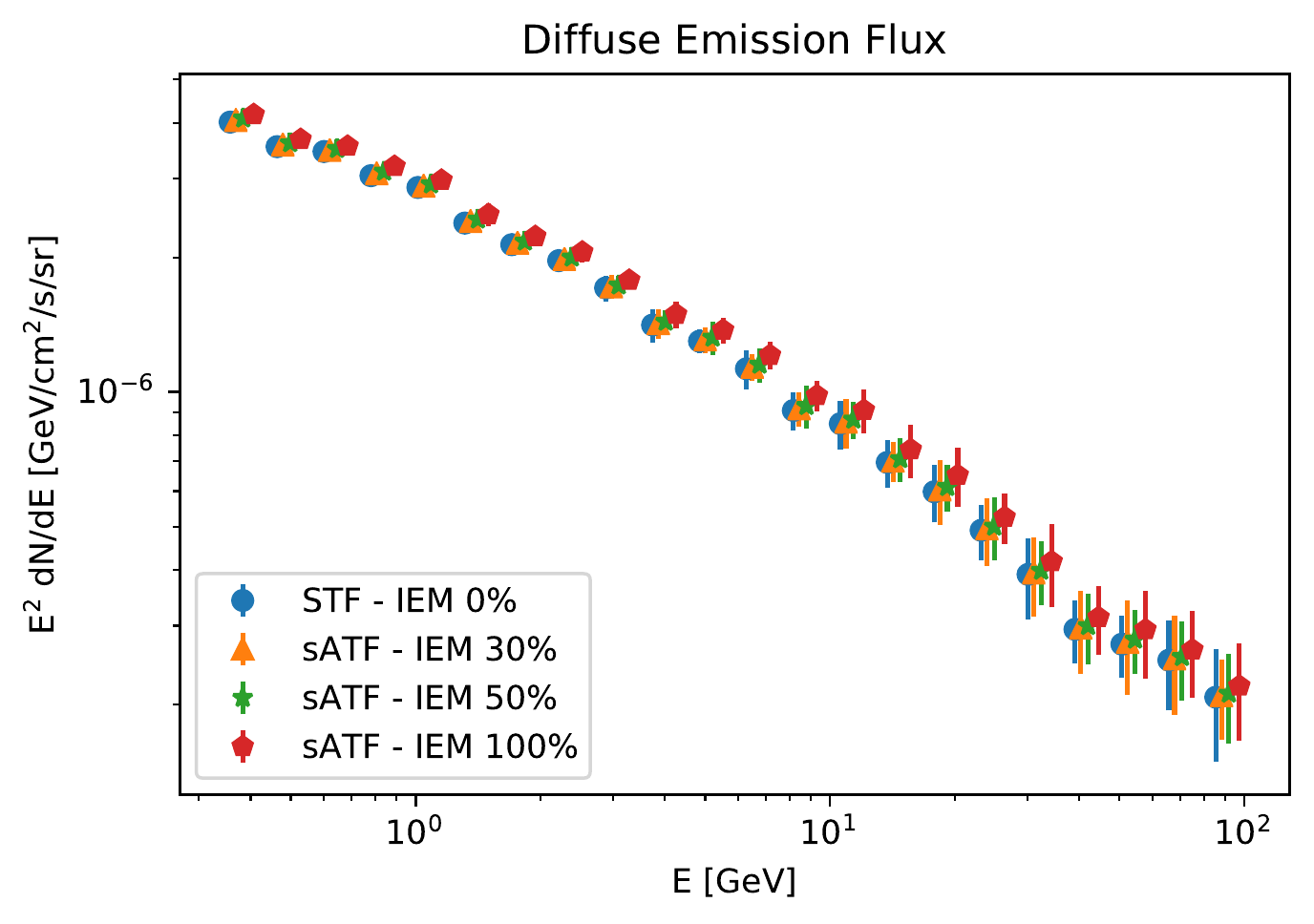}
\includegraphics[width = 0.45\textwidth]{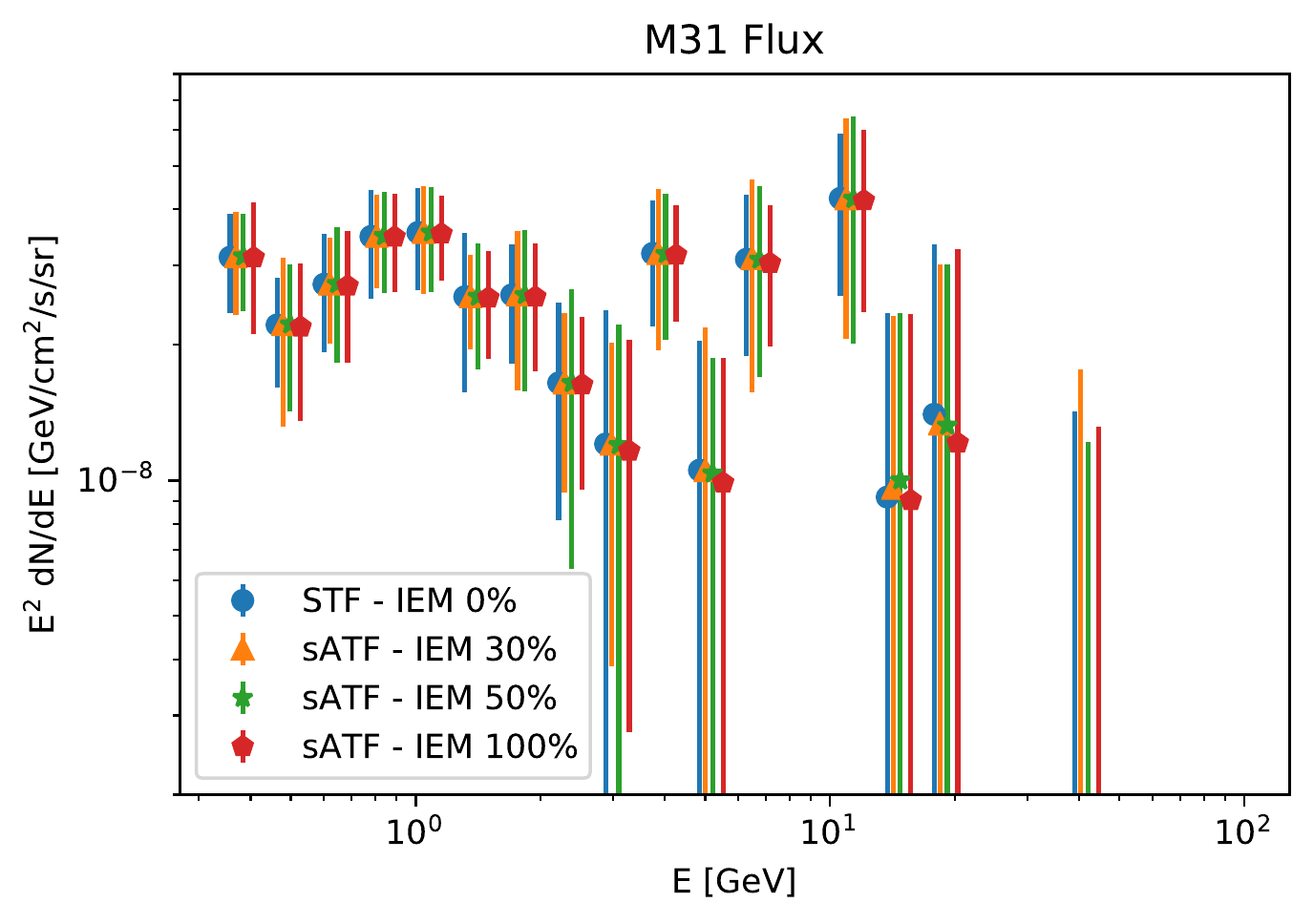}
\caption{Best-fit spectra of the IEM \textbf{(top)} and M31 \textbf{(bottom)} model components using four different baseline configurations, where we vary the freedom on the IEM spatial parameters: 0\% (STF), 30\%, 50\%, and 100\%. 
For more clarity, each point is slightly shifted on purpose on the energy scale with respect to the green points, placed at the true energies.} \label{fig:compa_spectrum} 
\end{figure}

For all three IEM configurations, the reconstructed spectra of the IEM and M31 model components result to be compatible with each other.   
In Fig.~\ref{fig:compa_spectrum}, we compare the IEM (top panel) and M31 (bottom panel) best-fit spectra obtained with the three sATF configurations and with the STF where no freedom on IEM spatial parameters is allowed. 
All spectra are extracted from the fits where M31 is described by a Gaussian template of $0.5\degree$ width. 
The reconstruction of the spectra for both the IEM and M31 model components is therefore robust against the variation allowed on the IEM spatial parameters, and compatible with STF results.

\begin{figure*}[p!]
\centering{
\includegraphics[angle=0,width=0.42\textwidth]{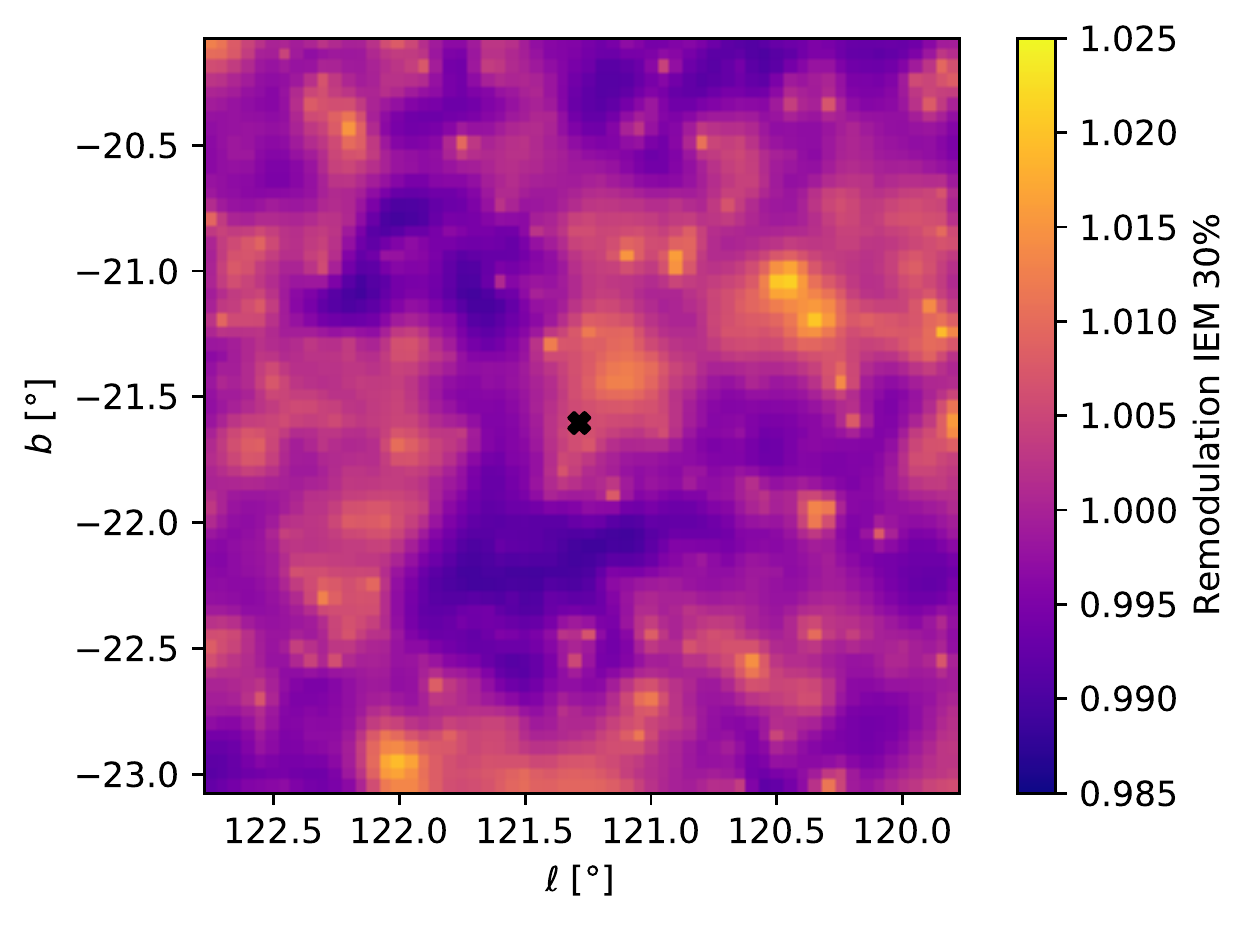}
\includegraphics[angle=0,width=0.42\textwidth]{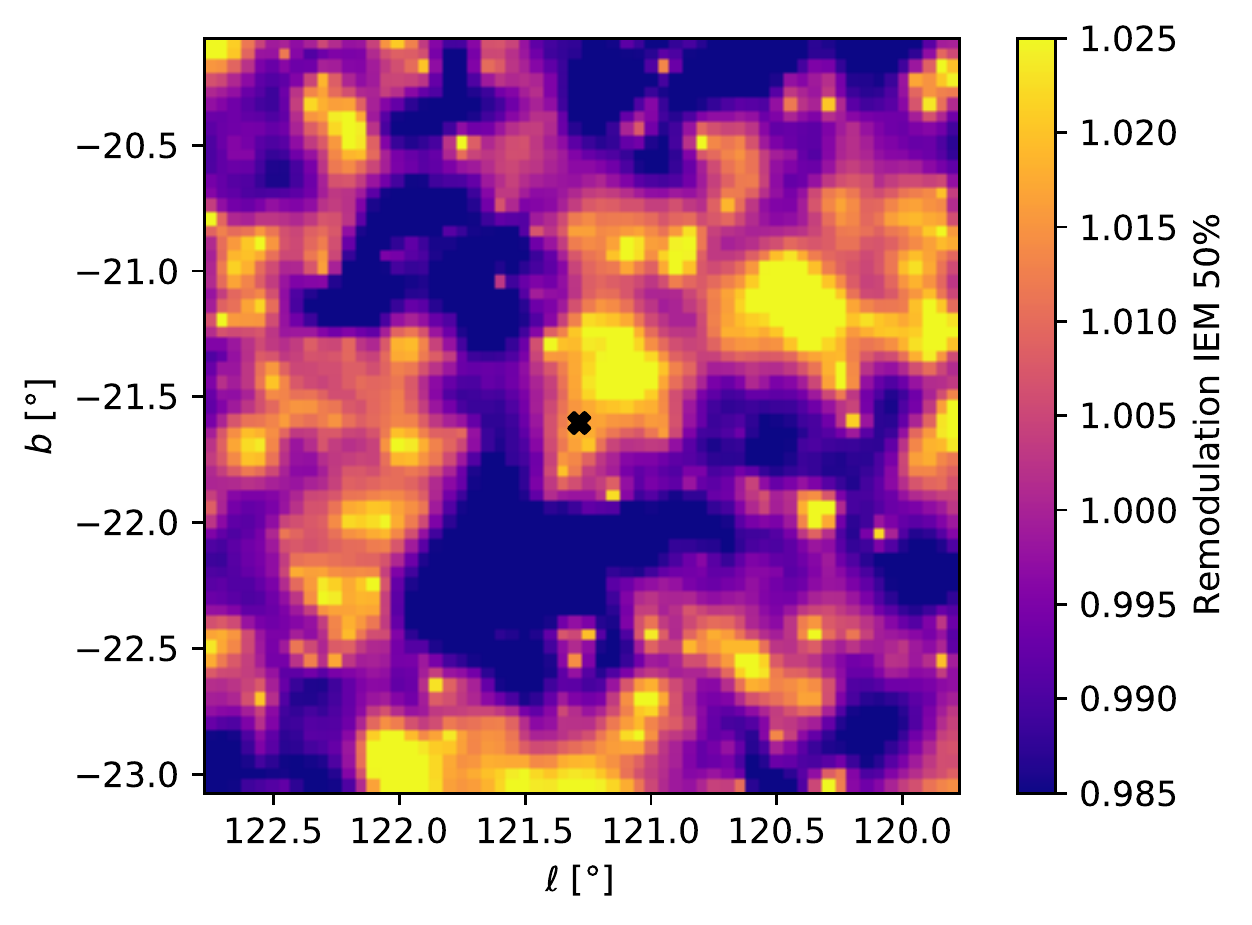}
\includegraphics[angle=0,width=0.42\textwidth]{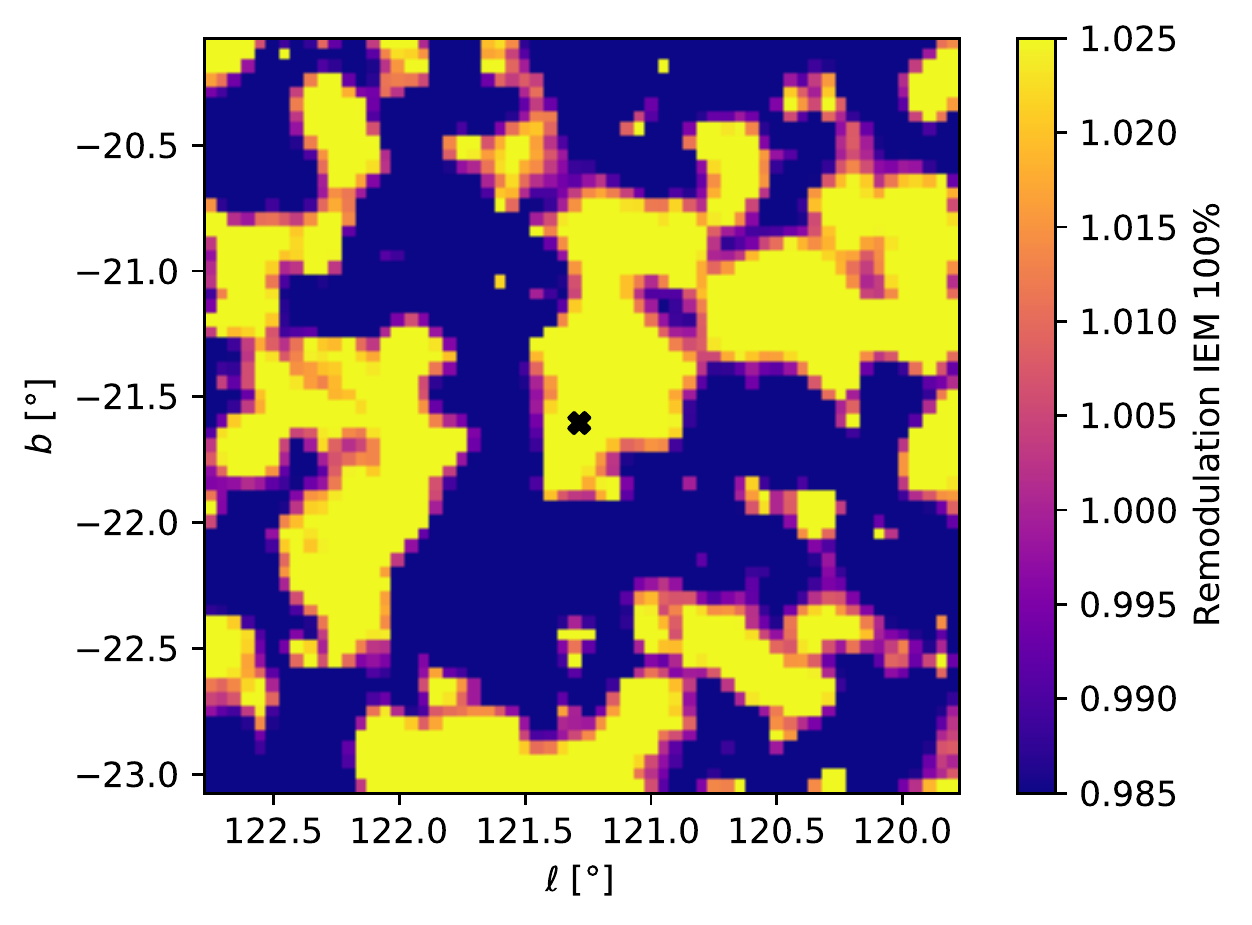}}
\caption{Remodulation map of the IEM component in the case of sATF for the Gaussian template ($0.5\degree$ width) for the IEM configurations 30\% \textbf{(top left)}, 50\% \textbf{(top right)}, and 100\% \textbf{(bottom)}. The black cross indicates the center of M31 given by the 4FGL.}
 \label{fig:remod_map_iem} 
\end{figure*}

In Fig.~\ref{fig:remod_map_iem}, we show the \textit{remodulation map} of the IEM component for the three IEM configurations, for a Gaussian template of $0.5\degree$ width.
The remodulation map indicates how much a component contributes to the total flux observed in each pixel, and what is its variation w.r.t. the input template. 
We notice that the IEM remodulation map for the 100\% configuration shows some significant re-modulation around the position of M31. 
This over-absorption is instead absent if we decrease the freedom on the IEM spatial parameters to 30\% and 50\%.
We find a negligible relative difference when comparing the outputs of the 50\% and 30\% configurations (0.02\% on average over the ROI, with 1\% variation for the pixel corresponding to M31 center).
Instead, the 100\% configuration remodulation map show variations at the \% level with respect to the 50\% and 30\% configurations (about 3\% on average over the ROI, with 5-6\% variation for the pixel corresponding to M31 center). We note that STF and sATF fits provide a reconstruction of the IEM component with a statistical uncertainty of 4-5\% at low energies, \textit{i.e.} the most constrained energies, that increases up to 28\% at the highest energies where there is less statistics.

Given that the best-fit extension angles and the reconstructed spectra are well compatible for the three IEM configurations, we adopt the IEM 30\% configuration for the rest of this study. 

\subsection{sATF with uniform spatial templates}

We perform a scan where M31 is modeled by a flat disk of radius extending from $0.025\degree$ (PS-limit) to 1\degree.
From the log-likelihood profile interpolation, we find two minima in this case: the first minimum gives an extension of $\theta = 0.40\degree\pm0.04$ which is compatible with the value of $0.33\degree \pm0.04$ found by \cite{DiMauro:2019frs} for a uniform disk template. 
The second minimum is located at $\theta = 0.95\degree\substack{+0.05 \\ -0.18}$. Both extensions have been detected with significance of 8.5$\sigma$.

The first minimum is compatible with the bulge size.
The second one can be instead explained by the presence of a more extended structure such as the stellar disk~\cite{Tamm_2012},
or a bubble-like component~\cite{Pshirkov:2016qhu}, 
while it does not seem to be consistent with the extent of the M31 disk as observed in optical light ($3.2\degree$)\footnote{NASA/IPAC EXTRAGALACTIC DATABASE - \url{https://http://ned.ipac.caltech. edu/}.}~\cite{Feng:2018vsl}.

Motivated by previous evidence for bubble-like structures lying above and below M31 galactic plane~\cite{Pshirkov:2016qhu}, we test a specific model for these structures. 
We perform the fits of nested models where we add an M31 bubble component to our previous best-fit models, the $0.5\degree$ width Gaussian bulge and the $0.4\degree$ radius uniform disk. We model the bubbles spectrum with a power law of index $-2.3$~ \cite{Pshirkov:2016qhu}. Our spatial template includes two uniform disks of $0.45\degree$ radius lying perpendicularly to the M31 galactic plane. The center of each bubble is offset by $0.45\degree$ from the center of M31.
We find only marginal evidence ($2.3\sigma$ at best) for the bubble component, 
a bit less significant than what found by~\cite{Pshirkov:2016qhu} who reported a $5.2\sigma$ evidence. 

\subsection{sATF with Einasto templates}
We finally test Einasto spatial templates as motivated by recent stellar mass models of M31.
In this case, we do not perform any scan over the Einasto template parameters since we want to test specific models tracing M31 stellar components: nucleus, bulge, and disk.
We keep the same spectral template of M31 extracted from the 4FGL for the nucleus and the bulge, while we use a power law of index -2.4 to model the spectrum of the disk~\cite{Pshirkov:2016qhu}.

\begin{figure}[h!]
\centering
\includegraphics[width=0.45\textwidth]{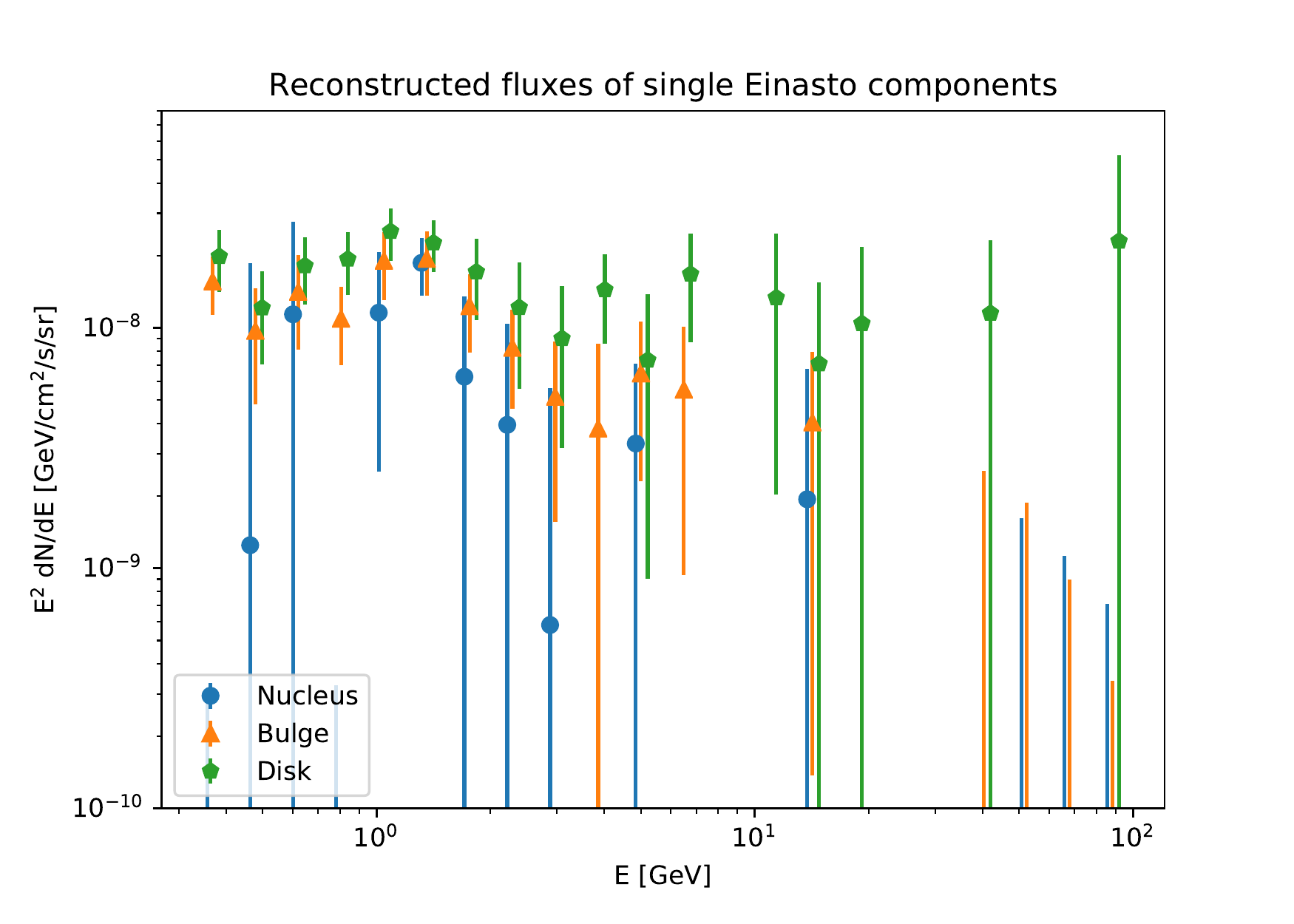}
\caption{Comparison of M31 reconstructed flux for the individual spatial Einasto models with either a nucleus, a bulge, or a disk modeling M31 emission.}
\label{fig:Compare_spectrum_Einasto_indiv} 
\end{figure}

We test the evidence for each of these three components, separately, on top of our baseline model with sATF (IEM 30\% configuration).
As for the results in Tab.~\ref{tab:sATF_Einasto}, the nucleus, bulge, and disk are all, individually, detected at about $7.5-8.5\sigma$.
In Fig.~\ref{fig:Compare_spectrum_Einasto_indiv}, we display the best-fit spectra of these three  templates when fitted as a single additional component. The reconstructed spectral fluxes of the nucleus and the bulge components follow the same trend with a bump between 1-2~GeV and a cut-off at higher energies. In the case of the disk model, a similar bump feature is also observed at low energies. In addition, the spectral reconstruction shows a high energy tail in the spectral flux. The authors of~\cite{Eckner:2017oul} report a similar observation which could potentially be confirmed by the next-generation of $\gamma$-ray experiments such as CTA~\cite{CTAConsortium:2018tzg} and LHAASO~\cite{Bai:2019khm}.

\begin{table}[ht!]
\centering{\textbf{\sffamily\bfseries  sATF results for Einasto spatial templates}}
\vspace{0.2cm}

\centering
\begin{tabular}{c|c|c}
\hline
Model &  $-2\ln \mathcal{L}$ & $\sigma_{\rm det}$  \\
\hline
\hline
 no M31 & 311107.65   & -   \\ 
\hline
N  &  311015.52  & 7.5  \\ 
B  &  311010.85  &  7.7  \\ 
D &   310997.55 &   8.4 \\  
\hline
N+B  &311010.06 &  7.8 \\ 
N+D & 310986.28   &  9.0  \\ 
B+D & 310986.16  & 9.0   \\ 
\hline
N+B+D &  310985.30 &  9.0  \\ 
\hline
\end{tabular}
\caption{Log-likelihood value, $-2\ln \mathcal{L}$, and detection significance, $\sigma_{\rm det}$, of the M31 model over the model without M31. We consider Einasto spatial templates for nucleus (N), bulge (B), and disk (D) and their combinations.}
\label{tab:sATF_Einasto}
\end{table}

We also run fits for all possible pair combinations and for the setup with all three templates in the fit.
We compute the test statistic and significance of all combinations of nested models to determine the significance of each component (Tab.~\ref{tab:TS_M31_einasto_nested}): \begin{itemize}
\item The addition of a disk to the nucleus (bulge) model improves the fit with a significance for the disk at $2.9\sigma$ ($2.4\sigma$). Adding instead a nucleus or a bulge component to the disk model shows a very small improvement of the fit with a significance of less than $1\sigma$.
\item The addition of a bulge to the nucleus model does not improve significantly the fit ($0.2\sigma$). The addition of a nucleus to the bulge does not show any improvement either. 
\item The case of the fit with all three sub-components (nucleus, bulge, and disk) does not provide any significant improvement over the nucleus+disk and bulge+disk models. \end{itemize}

The preference for a disk-like extended structure is in agreement with our previous findings for the uniform spatial disk template. 
The extension of the Einasto disk (1.08$\degree$ semi-major axis) is indeed compatible with the second minimum found in the uniform disk scan ($0.95\degree$).

\begin{table}[ht!]
\centering{\textbf{\sffamily\bfseries Einasto spatial templates: nested models comparison}}
\vspace{0.2cm}

\begin{tabular}{c|c|c|c}\hline \hline
Model & $\sigma$ & Model & $\sigma$ \\

\hline

N+B vs N   &  0.2 &     N+D vs D & 0.8\\[0.1cm]

N+D vs N & 2.9 &      B+D vs D & 0.8  \\[0.1cm]

N+B+D vs N & 3.0  &     N+B+D vs D & 0.9 \\

\hline

N+B vs B &  $\sim0$  &N+B+D vs N+B &  2.4\\[0.1cm]

B+D vs B &  2.4 & N+B+D vs N+D & $\sim 0$  \\[0.1cm]

N+B+D vs B & 2.5 & N+B+D vs B+D &  $\sim 0$ \\

\hline
\end{tabular}
\caption{Significance, $\sigma$, of the nested models considering combinations of nucleus (N), bulge (B), and disk (D) templates.}
\label{tab:TS_M31_einasto_nested}
\end{table}

\subsection{Akaike Information Criterion (AIC) model comparison}
In order to properly determine which model best describes the spatial distribution of M31, we compare our different non-nested templates using the AIC score, as defined in Appendix~\ref{sec:AIC}. 

In Tab.~\ref{tab:AIC_Compare_model}, we quote the AIC differences between the various template combinations. 
Based on the $\Delta$AIC values, we can rank for the very first time the different models used to describe the morphology of M31: the lower the AIC score, the better the model is.
In general, models for which the $\Delta$AIC relative to the model with a lower AIC is less than 2 are typically  considered to have substantial support from data~\cite{Burnham_1998}.
The uniform templates with $\theta = 0.4\degree \pm 0.04$ and $\theta = 0.95\degree\substack{+0.05 \\ -0.18}$ turn out to be the best models we tested with an almost identical score (AIC$_{\rm min}$), followed by the nucleus+disk, the bulge+disk and disk-only Einasto templates, and finally by the Gaussian template of width $\theta = 0.48\pm 0.07\degree$.

We note that, from the log-likelihood results only (\textit{i.e.} nested model comparison), the addition of a bulge or a nucleus to the disk model improves the likelihood but not enough ($0.8\sigma$) to claim the presence of an additional component. The AIC score, on the other hand, factors in the number of effective parameters of the fits. In the case of the bulge+disk model, the improvement in the log-likelihood is not enough to counterbalance the increase of effective parameters due to the addition of the bulge component. The nucleus+disk model, on the other hand, has about the same number of effective parameters -- due to the small size of the nucleus --, while the log-likelihood improves. Thus, for the case nucleus+disk, we get a significantly better model.

\begin{table*}[ht!]
\centering{ \textbf{ $\Delta${\sffamily\bfseries AIC model comparison} }}
\vspace{0.2cm}

\centering
\begin{tabular}{c|c|c|c|c|c}
\hline
 Model & Gaussian $0.5\degree$ & Uniform disk $0.4\degree$ & Uniform disk $0.9\degree$ & Einasto D & Einasto B+D \\
\hline
\hline

Uniform $0.4\degree$ &  -56.51  & -& - & - & -  \\[0.1cm]

Uniform $0.9\degree$  & -55.93  & 0.58 & - & - & - \\[0.1cm]

Einasto D & -44.79 & 11.71 & 11.13 & - & - \\[0.1cm] 	

Einasto B+D & -44.67 & 11.26 & 21.65 & 0.12 & - \\[0.1cm] 			

Einasto N+D & -55.33 & 1.18 & 0.60 & -10.53 & -10.66 \\[0.1cm]

\hline
\end{tabular}
\caption{$\Delta$AIC computed for (Row - Column). A positive (negative) $\Delta$AIC indicates that the model listed in the corresponding column gives a better (worse) fit to data than the model in the corresponding row. }
\label{tab:AIC_Compare_model}
\end{table*}
\section{Image reconstruction of M31 morphology}\label{sec:atf}

\begin{figure}[h!]
\centering
\includegraphics[scale=0.65]{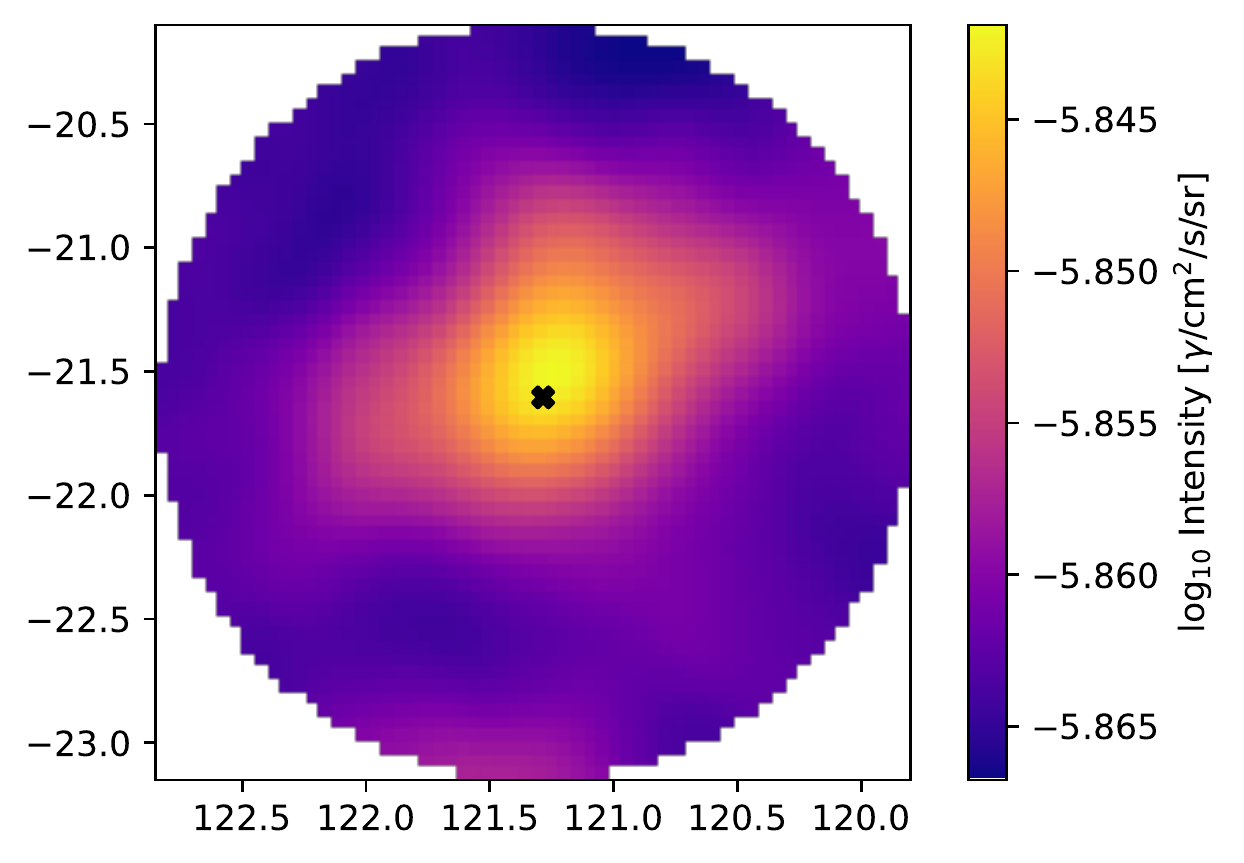}
\caption{Reconstruction of the $\gamma$-ray emission of M31 using the full adaptive template fitting (ATF) technique. 
The black marker indicates the position of the source given by the 4FGL at ($l=121.285\degree$, $b=-21.604\degree$)~\cite{Fermi-LAT:2019yla}.
} 
\label{fig:ATF_free_morpho} 
\end{figure}

We finally take full advantage of \SF's flexibility by performing an adaptive template fit allowing a 100\% freedom on the spatial and spectral parameters of the M31 component.
We keep the same input M31 spectral shape from the 4FGL as above. As for the spatial part, we do not assume any spatial input morphology, but we rather define a large circular region of $1.5\degree$ radius around the source position, within which the M31 component is adjusted to the data in each pixel and each energy bin.
The reconstructed morphology of M31 is shown in Fig.~\ref{fig:ATF_free_morpho}.

\begin{figure}[h!]
\centering
\includegraphics[scale=0.6]{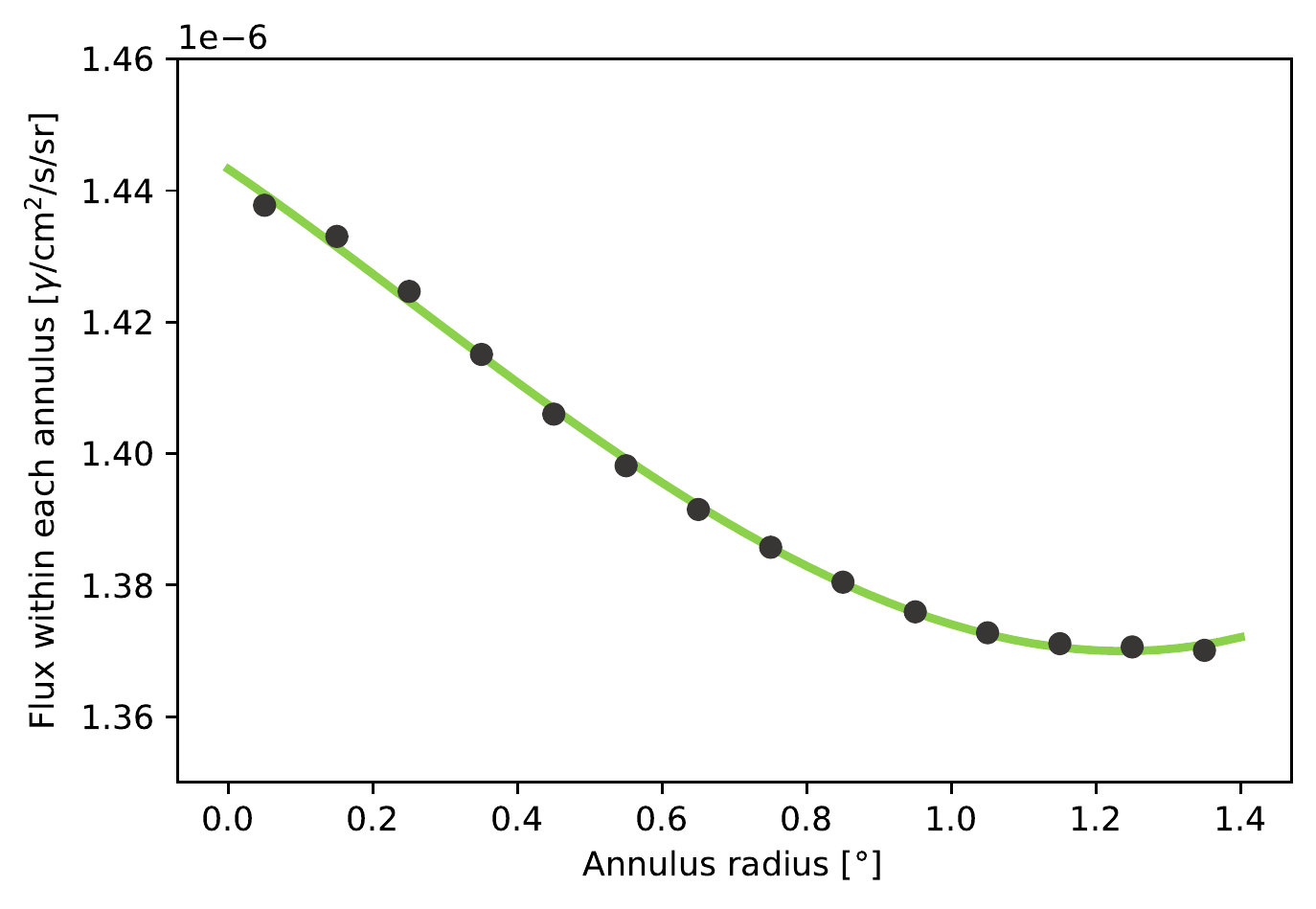}
\caption{Intensity profile of M31 built using concentric annulii of width $0.1\degree$ (dots). The green line shows the total fit of the intensity profile which corresponds to the combination of a Gaussian function and a power law.}
\label{fig:intensity_profile}  
\end{figure}

From the reconstructed M31 image,  we build the intensity profile of M31 shown in Fig.~\ref{fig:intensity_profile} by summing the contributions of all pixels in concentric annuli spaced of $0.1\degree$ from $0\degree$ to $1.5\degree$. The intensity of each annulus is then averaged over the solid angle $\Delta\Omega_{\rm{annulus}} = N \Delta\Omega_{\rm{pixel}}$, where $N$ is the number of pixels in a given annulus.
The annulii are centered around M31 4FGL position.

The intensity profile appears rather flat and does not follow a simple known function. We fit the intensity using a combination of a Gaussian function and a power law given by:
\begin{equation}
\Xi(x) = a \exp\left\{ \frac{-(x-\mu)^2}{2 \sigma^2}\right\}  + b x^c,
\end{equation}
where the best-fit coefficients are found to be  $a=1.48\times 10^{-6}$, $\mu = -1.04 \times 10^{-1}$, $\sigma = 4.35$, $b=4.72 \times 10^{-8}$, and $c=2.37$.

We estimate the goodness of fit using the $R^2$ regression sum of squares method, defined in Appendix~\ref{sec:Rgoodness}. 
We find a value of $R^2 = 0.9981$ which indicates a very good fit of the intensity by our model $\Xi(x)$.

These two functions in $\Xi(x)$ can be interpreted as associated to the two distinct components of M31 emission also hinted by our previous studies: the Gaussian function would model M31 central emission as it is peaked at the center of the source and decreases at larger radii, while the power law could correspond to the extended structures around the galaxy, such as galactic bubbles, the outer part of the disk, and/or a larger cosmic-ray halo. We find an inflection point of the intensity profile at $0.3\degree$ by taking its second derivative. This corresponds to the angle at which the power law starts kicking in and may delimit the apparent size of the bulge in $\gamma$ rays. This value is compatible with the result of~\cite{DiMauro:2019frs} who found an extension angle of $0.33\degree \pm0.04$ using a uniform disk template to model M31's bulge, and also marginally consistent with our first minimum of the uniform disk scan $0.40\degree \pm0.04$.

\section{Discussion and conclusions}
\label{sec:conclusion}
In the present work, we analyzed the \Fermi-LAT $\gamma$-ray emission at GeV energies from the direction of M31.
We re-assessed the evidence for an additional {\it extended} $\gamma$-ray source at the position of M31, and carefully characterized its morphology by quantifying systematic uncertainties from the IEM foreground emission. 
To this end, we introduced the following methodological novelties: 
\begin{itemize}
\item For the first time, we adapted and applied the \SF~code to the analysis of extended $\gamma$-ray sources, generalizing its scope (which was previously only limited to the inner Galaxy analysis).

\item We fully exploited the power of \SF~of accounting for a large number of free parameters by defining three setups which allowed us (a) to recover results from the literature and estimate the systematic uncertainties due to the PSF mis-modeling (STF, Sec.~\ref{sec:std}), (b) to properly compare non-nested models with various spatial templates, (c) to quantify systematic uncertainties from IEM mis-modeling (sATF, Sec.~\ref{sec:semiATF}), and (d) to reconstruct the {\it image} of M31 in a template-independent way (ATF, Sec.~\ref{sec:atf}).

\item We tested a large number of M31 spatial templates of three kinds, the Gaussian and uniform models motivated by previous literature, and the Einasto spatial templates which are directly motivated by the different M31 stellar components (nucleus, bulge, and disk).

\item For the first time, we ran a proper model comparison of our different spatial models for the emission of M31, and ranked them based on their AIC score. This allowed us to be able to state what morphology is preferred by the data, among the ones we tested. 
\end{itemize}

We summarize below our main findings: 
\begin{itemize}
\item We confirm the evidence for an {\it extended} $\gamma$-ray emission from the direction of M31 ($\sim 9\sigma$ detection significance). 
\item For a Gaussian spatial template, the best-fit angular extension is $\theta = 0.48\degree \pm{0.07}$ (68.27\% statistical uncertainty). 
This determination is very robust against systematic uncertainties related to mis-modeling of the PSF and of the IEM. 
We quantify PSF systematic uncertainties at the level of 1.5\%. To estimate IEM-related uncertainties, we set up sATF runs with increasing freedom in the variation of the IEM spatial parameters (30\%, 50\% and 100\% allowed variation).
We quantify the IEM systematic error to be 2.9\%.

\item When modeling M31's morphology with a uniform spatial template, we interestingly find two minima of the log-likelihood profile: the first one at an angular extension of $0.4\degree \pm0.04$ associated to M31 bulge, and the second at $0.95\substack{+0.05 \\ -0.18}\degree$ indicating the presence of a more extended structure around M31, such as the galactic bubbles~\cite{Pshirkov:2016qhu}.
We explicitly tested the presence of bubbles-like structures lying above and below the M31 galactic plane on top of a centrally concentrated emission. We only find marginal evidence ($< 3\sigma$) for this additional emission component.

\item The use of spatial templates tracing the stellar distribution of M31 nucleus, bulge, and disk components provides some more insight on the physical nature of the $\gamma$-ray emission.
We observe that the data prefer to complement centrally concentrated components (nucleus- or bulge-only models) with an extended disk-like component ($\sim 3\sigma$ evidence), which has an angular extension of about $1\degree$ (semi-major axis).
This is a further indication of the fact that the M31 $\gamma$-ray emission is indeed more extended 
than what was previously found in most of the literature showing an evidence for the bulge only component~\cite{DiMauro:2019frs, Fermi_M31_M33, Feng:2018vsl}.
On the other hand, we cannot claim evidence for a centrally concentrated component (nucleus or bulge) on top of a disk-only model, as we found a marginal improvement of $0.8\sigma$ only.
The case of a fit with all three sub-components (nucleus, bulge, and disk) instead does not provide any significant improvement either over the nucleus+disk or
bulge+disk models.

\item We properly compare spatial template models through the evaluation of the AIC score of each model. Based on $\Delta$AIC values, we find the data rather prefer flat templates to describe the $\gamma$-ray emission of M31. As a matter of fact, the uniform disk templates come first in our ranking, followed by the Einasto nucleus+disk model, then the Einasto disk and bulge+disk models, and finally the Gaussian templates. 
Again, this highlights the preference for a more extended emission, which goes beyond M31's bulge. 

\item We find that the M31's reconstructed spectrum is extremely robust against all variations of fit setups and spatial templates we tested. The emission from M31 is significant up to, at most, 10 GeV in energy.

\item We reconstruct the morphology of M31 in a fully template-independent way. To this end, we do not impose any a priori spatial distribution for M31, and we leave the \SF~spatial parameters to freely adjust to the data. 
From the reconstructed angular intensity profile, we show that the M31 intensity profile can be fitted by a two-component parametrization made up by a Gaussian function, dominant within $\sim 0.3\degree$, and a power law at larger radii. The latter component could correspond to a superposition of several contributions such as the outer part of the disk, galactic bubbles similar to the ones of the Milky Way, and/or a large comic-ray halo surrounding M31~\cite{Do:2020xli}.
\end{itemize}

Our results therefore indicate that the emission from M31 extends beyond its nucleus and bulge, with significant emission up to at least $1\degree$.
This can be compatible either with a flat disk-only emission, or with a two-component model such as a nucleus+disk. 
We stress that these components should trace the stellar distribution in M31, rather than gas regions given the lack of evidence for $\gamma$-ray gas-correlated emission~\cite{Ackermann_2017_M31_M33}. 
A stellar origin of M31's extended emission is also strongly supported by the reconstructed spectral energy distribution, which is not compatible with the Milky Way IEM spectrum and clearly distinguishable from  it.
A plausible explanation for the emission can be, for example, a large population of millisecond pulsars in the disk and bulge of M31~\cite{Eckner:2017oul}.
This finding is of relevance especially for models trying to interpret, in a common framework, the $\gamma$-ray emission from M31 and from the Milky Way inner region, a.k.a.~the GeV excess~\cite{Feng:2018vsl,Eckner:2017oul}.

With its better resolution and wide energy coverage, the upcoming $\gamma$-ray Cherenkov telescopes CTA~\cite{CTAConsortium:2018tzg} and LHAASO~\cite{Bai:2019khm} will contribute to demystifying the nature of M31's emission, by, for instance, testing the high-energy tail of its spectrum and possibly detecting an interstellar emission component.

\bigskip

\begin{acknowledgments}
We would like to thank P.~D.~Serpico, C.~Eckner, F.~Donato, and M.~Di Mauro for useful and enlightening discussions.
We also thank C.~Eckner for comments on the manuscript, and C.~Weniger for initial support with the \SF~code.
We acknowledge support from the ANR JCJC 2019 grant ``GECO'' (PI: F.~Calore).
This work has been done thanks to the interactive servers of the Univ. Savoie Mont Blanc - CNRS/IN2P3 MUST computing center. We would also like to thank M. Gauthier-Lafaye for his work on the servers upon our multiple requests.
\end{acknowledgments}

\begin{appendix}

\section{Description of the fitting algorithm \SF}\label{App:SF}

For each emission model component of the fit, the $\gamma$-ray flux is modeled for a pixel $p$ and an energy bin $b$ by the following equation:
\begin{equation}
\Phi_{pb}  = \sum_k T_p^{(k)} \tau_p^{(k)} \cdot S_b^{(k)} \sigma_b^{(k)} \cdot \nu^{(k)}
\end{equation}
where $T_p^{(k)}$ and $S_b^{(k)}$ are the spatial and spectral templates respectively of the input component $k$ and $\tau_p^{(k)}$ and $\sigma_b^{(k)}$ are their associated spatial and spectral modulation (or nuisance) parameters. \footnote{As for point-like sources, they are described only by a spectral template.}
The quantity $\nu^{(k)}$ represents an overall normalization factor. More details can be found in~\cite{Storm:2017arh}. 

The modeled $\gamma$-ray flux is then divided by the \Fermi-LAT exposure and convolved with the \Fermi-LAT PSF to derive the expected number of $\gamma$ rays per energy bin and per pixel. This model is then fitted to the \Fermi-LAT count map, minimizing the following total log-likelihood:
\begin{equation}
\ln \mathcal{L} = \ln \mathcal{L}_P + \ln \mathcal{L}_R \, , 
\end{equation}
where $\mathcal{L}_P$ is the standard Poisson likelihood, and $\mathcal{L}_R$ represents the regularization term which controls the variation of the modulation parameters.

The optimization is performed with the L-BFGS-B (Limited memory BFGS with Bound constraints) algorithm \cite{Zhu_1997, Bfgs_b, bfgs_2},
which is similar to the BFGS quasi-Newtonian method but uses less memory. 
These algorithms are developed to find the local extrema of functions based on Newton's method where a second degree approximation is used to find the minimum function. They both use the Hessian inverse matrix estimation to search for the parameters maximizing the function.
The advantage of the L-BFGS-B is that the algorithm does not store a dense $n \times n$ approximation to the inverse Hessian matrix but instead stores a few vectors representing the approximation implicitly. Therefore, the L-BFGS-B algorithm is well suited for very large-scale problems, \textit{i.e.} with a large number of parameters.
In addition, the L-BFGS-B algorithm supports boundary conditions which is necessary to impose non-negativity constraints on the parameters, see discussion in~\cite{Storm:2017arh}.

The uncertainties of individual model parameters and component fluxes are computed using a sampling method.
This technique circumvents the computation of the covariance matrix which would require a significant computational time due to the large number of parameters. 
It relies on the Fisher information matrix given by:
\begin{equation}
\mathcal{I}(\theta)_{ij} = - \bigg \langle \frac{\partial^2}{\partial \theta_i \partial \theta_j}  \ln \mathcal{L} \bigg \rangle_{\mathcal{D}(\theta)},
\end{equation} 
where $\theta$ are the model parameters and $i,j$ define a matrix element. The average is calculated using mock data generated from the best-fit model $\mathcal{D}(\theta)$. 
The Fisher matrix is then decomposed into triangular and diagonal matrices using sparse Cholesky decomposition \cite{Cholesky} from which sample model parameter vectors $\delta \theta$ are computed. Mean values of component fluxes and model parameters are derived from the best-fit value of $\theta$ while standard deviations are derived by computing model predictions for $\theta +\delta \theta$ which are averaged over many samples.

\section{Statistical analysis framework}\label{stats}

The total likelihood given by \SF~is the sum of a Poisson likelihood $\mathcal{L}_P$ and a regularization term  $\mathcal{L}_R$ that controls the modulation parameters.
To probe the extension of M31, we use the likelihood ratio (LR) test statistic where we state two hypotheses.
The null hypothesis $\mathcal{H}_0$ corresponds to the absence of M31, where only the diffuse components (IEM, IGRB) and PS are considered, while the alternative hypothesis $\mathcal{H}_1$ includes an emission component for M31. The LR test statistic (TS) is defined as:
\begin{equation}
\rm{TS} = -2 \ln \frac{\mathcal{L}_{\mathcal{H}_0}( \boldsymbol{\hat{\theta}}_{\rm{Diffuse}},  \boldsymbol{\hat{\theta}}_{\rm{IGRB}},  \boldsymbol{\hat{\theta}}_{\rm{PS}})}{\mathcal{L}_{\mathcal{H}_1}(\boldsymbol{\hat{\theta}}_{\rm{M31}}, \boldsymbol{\hat{\theta}}_{\rm{Diffuse}},  \boldsymbol{\hat{\theta}}_{\rm{IGRB}},  \boldsymbol{\hat{\theta}}_{\rm{PS}})},
\end{equation}
where $\boldsymbol{\theta}_{\rm{M31}}$, $\boldsymbol{\theta}_{\rm{Diffuse}}$,  $\boldsymbol{\theta}_{\rm{IGRB}}$,  and $\boldsymbol{\theta}_{\rm{PS}}$ are the parameters corresponding to M31, IEM, IGRB, and PS components respectively. The hat refers to the values maximizing the likelihood functions $\mathcal{L}_{\mathcal{H}_0}$ and $\mathcal{L}_{\mathcal{H}_1}$.
The parameters $\boldsymbol{\theta}_i$ are all constrained in $\mathbb{R}^+$.
Typically, the LR TS is distributed as a $\chi^2_q$ distribution if the null hypothesis $\mathcal{H}_0$ is true, according to Wilk's theorem~\cite{wilks1938}, where $q$ is the number of free parameters. However, in this case, the parameters are not allowed to take on negative values. As a result, the Wilk's theorem does not hold anymore since the values of the parameters of $\mathcal{L}_{\mathcal{H}_0}$ are on the boundary of the allowed parameter space. 
For this theorem to be valid, the maximum likelihood estimators must find a non-boundary value where the function is differentiable~\cite{gascuel2005mathematics}.  
This problem can be solved using the Chernoff theorem \cite{chernoff1954} for nested models which combines a $\chi^2$ and a Dirac function $\delta$ at 0: 

\begin{equation}P(TS) = 2^{-n} \left[\delta(0) + \sum^n_{i=1} \binom{n}{i} \chi^2_i (TS) \right] \, , 
\end{equation}
where the $2^{-n}$ term is the number of distinct ways that $n$ energy bins can take on non-negative values.
The $\delta$ function ensures that all energy bins fulfil the non negativity condition
while the binomial coefficient $\binom{n}{i}$ describes the number of possible configurations of non-negative amplitudes
where each configuration has a $\chi^2_i$ distribution \cite{Macias:2016nev}.

The TS can then be interpreted in terms of standard deviations $\sigma$. First, one needs to compute the $p$-value, also known as the \textit{survival function}, $p = P( \rm{TS}> \Lambda) = 1 - \rm{CDF}$, where CDF is the cumulative distribution. The survival function corresponds to the probability of obtaining a TS value larger than the given value $\Lambda$. In other words, it represents the proportion of data ``surviving'' above a certain value. 
The number of $\sigma$ is then derived by performing the square root of the inverse survival function ``InverseCDF'' of a $\chi^2_q$ distribution with argument $p$ and reads as:
\begin{equation}
\sigma = \sqrt{\rm{InverseCDF} (\chi^2_q, \rm{CDF}[p(\rm{TS}),\widehat{TS}] )},
\label{eq:sigma_skyfact}
\end{equation}
where $\widehat{\rm{TS}}$ is the observed TS value \cite{Macias:2016nev, Bartels:2017vsx}.
In our hypothesis testing (\textit{i.e.}~presence of M31), 
when the M31 spectral parameters are fixed, the number of free parameters $q$ is one, corresponding to the overall normalization $\nu^{\rm M31}$. However, when the spectral parameters are free to vary, the number of free parameters equals the number of energy bins~\cite{Bartels:2017vsx}.

\vspace{0.3cm}

\section{Non nested model comparison} \label{sec:AIC}
Comparing complex, non-nested, models is quite subtle since the number of effective parameters vary from a model to another. The larger the number of effective parameters, the easier it is for a fit to converge. Therefore, one needs to take into account the number of effective parameters in order to assess the goodness of the models from one to another.
This is what is required to meaningfully compare different spatial models for M31. 

In the case of non nested models, the $\delta$-$\chi^2_q$ mixture cannot be applied. 
As an alternative, model comparison can be performed using the Akaike Information Criterion (AIC) defined as \cite{AIC_1974}:
\begin{equation}
{\displaystyle \mathrm {AIC} = 2N_{\rm{Param}}^{\rm{eff}}-2\ln( {\hat {\mathcal{L}}_{\rm{Data}} } )},
\end{equation}
where $N_{\rm{Param}}^{\rm{eff}}$ is the number of estimated effective parameters in the model and $-2\ln({\hat {\mathcal{L}}^P}_{\rm{Data}})$ is the minimized Poisson likelihood function on the data. 

The effective number of parameters of each model is derived using \cite{Storm:2017arh}:
\begin{equation}
N_{\rm{Param}}^{\rm{eff}} = N_{\rm{Data}}^{\rm{eff}} - N^{\rm{eff}}_{\rm{DOF}},
\end{equation}
where $N_{\rm{Data}}^{\rm{eff}}$ is the effective number of data bins and $N^{\rm{eff}}_{\rm{DOF}}$ is the effective degree of freedom (DOF). 
The quantity $N_{\rm{Data}}^{\rm{eff}}$ is derived by averaging Poisson likelihood values of several sets of mock data obtained without refitting the data:
\begin{equation}
N_{\rm{Data}}^{\rm{eff}} = \langle -2 \ln \mathcal{L}_P(\theta)\rangle_{\rm{Mock}}.
\end{equation}
The second quantity $N^{\rm{eff}}_{\rm{DOF}}$, however, is defined by the average of the Poisson likelihood values obtained by refitting the mock data:
\begin{equation}
N^{\rm{eff}}_{\rm{DOF}} = \langle -2 \ln \mathcal{L}_P(\hat{\theta})\rangle_{\rm{Mock}},
\end{equation}
where $\hat{\theta}$ are the values of the model parameters that minimize the likelihood function.
In what follows, we use twenty sets of mock data to derive $N_{\rm{Data}}^{\rm{eff}}$ and $N^{\rm{eff}}_{\rm{DOF}}$.

\vspace{0.3cm}

\section{$R^2$ regression} \label{sec:Rgoodness}
In Sec.~\ref{sec:atf}, we use the $R^2$ regression sum of squares method to estimate the goodness of the intensity profile fit. 
The  $R^2$ regression is defined by~\cite{darlington2016regression}:
\begin{equation}
R^2 = \frac{SS_{\rm{Regression}} }{SS_{\rm{Total}}} = 1 - \frac{SS_{\rm{Residual}} }{SS_{\rm{Total}}},
\end{equation}
where 
\begin{align}
&SS_{\rm{Regression}} = \sum^N_{i=1} (\hat{Y}_i - \bar{Y})^2\\
&SS_{\rm{Residual}} = \sum^N_{i=1} (Y_i - \hat{Y}_i)^2\\
&SS_{\rm{Total}} = \sum^N_{i=1} (Y_i - \bar{Y})^2
\end{align}
where $Y_i$ and $\hat{Y}_i$ are the observed and best-fit intensity values in the annulus $i$ respectively and $\bar{Y}$ is the mean intensity. The term $SS_{\rm{Regression}}$ quantifies how much the best fit deviates from $\bar{Y}$, while $SS_{\rm{Residual}}$ estimates the discrepancy between the model and the actual intensity, and $SS_{\rm{Total}}$ measures the variability of the data.
The closer $SS_{\rm{Regression}}$ is to $SS_{\rm{Total}}$, the better the model, in contrast with $SS_{\rm{Residual}}$ where the closer the value to $SS_{\rm{Total}}$, the worse the model. The $R^2$ value is given between 0 and 1, 0 indicating a bad fit and 1 a perfect fit.

\end{appendix}

\bibliography{m31_biblio}

\end{document}